\documentclass[twocolumn,times,tighten]{aastex63}
\usepackage{natbib}

\accepted{by ApJ, 2023 December 5}
\shorttitle{Interstellar 3.2-3.3 Microns Absorption Feature} \shortauthors{Bernstein \& Geballe}
\begin{document}

\title{Analysis of the 3.2-3.3 $\mu$m Interstellar Absorption Feature on Three Milky Way Sightlines }


\author{L. S. Bernstein}
\affiliation{63 Forest Glen Ln., Topsham, ME 04086, USA, spectral1949@gmail.com}

\author{T. R. Geballe} \affiliation{Gemini Observatory/NSF's NOIRLab, Hilo, 670 N. A'ohoku Pl., HI 96720 USA}
\%================================================================
\vspace{4mm}
\begin{abstract}
  We report new analyses of spectra of the $3.2-3.3~\mu$m absorption feature observed in the diffuse interstellar medium toward three Milky Way sources: 2MASS $J17470898-2829561$ (2M1747) and the Quintuplet Cluster, both located in the Galactic center, and Cygnus OB2-12.  The $3.2-3.3~\mu$m interval coincides with the CH-stretching region for compact polycyclic aromatic hydrocarbons (PAHs).  We focus on the 2M1747 spectrum.  Its published optical depth spectrum contains residual telluric transmission features, which arise from the 0.06 difference in mean airmasses between the observations of the source and its telluric standard star.  We corrected the published  spectrum by adding the airmass residual optical depth spectrum.  The corrected spectrum is well fit by a superposition of four Gaussians.  The absorption spectra of the other two sources were also fit by four Gaussians, with similar central wavelengths, widths, and relative peak opacities.  We associate the three longer wavelength Gaussians covering the $3.23-3.31~\mu$m interval with compact PAHs in positive, neutral, and negative charge states.  We identify the shortest wavelength Gaussian, near 3.21 $\mu$m, with irregularly-shaped PAHs.   Constraints imposed by spectral smoothness on the corrected 2M1747 spectrum, augmented by a PAH cluster formation model for post-asymptotic giant branch stars, suggests that $> 99$\%\ of the PAHs in the diffuse interstellar medium reside in small clusters.  This study supports the PAH hypothesis, and suggests that a family of primarily compact PAHs with a C$_{66}$H$_{20}$ (circumvalene) parent is consistent with the observed mid-infrared and ultraviolet interstellar absorption spectrum.
  
\end{abstract}
\keywords{Interstellar dust (836); Interstellar dust extinction (837); Interstellar dust processes (838); Carbonaceous grains (201); Polycyclic aromatic hydrocarbons (1280); Infrared astronomy (786); Molecular spectroscopy (2095); Astrochemistry (75); Ultraviolet astronomy (173); Post-asymptotic giant branch stars (2121)}

\section{INTRODUCTION}

  The precise identities of the presumed large molecules that produce the hundreds of narrow diffuse interstellar absorption bands (DIBs) at optical and infrared (IR) wavelengths, as well as  the generally  broader emission and absorption bands at IR wavelengths, remain a mystery.  The carriers of these spectra are generally attributed to large hydrocarbons in gaseous or solid form \citep[see the reviews by][]{her95, tie95, ful00, sar06, tie08, cam14}.  The observational data  encompass numerous terrestrial and satellite-based spectra spanning the microwave through the ultraviolet spectral range, from a multitude of galactic and extragalactic sources.  To date, only three large molecules have been clearly identified as accounting for a few of the spectral features, the fullerenes C$_{60},$ C$_{60}^{+}$ , and C$_{70}$ \citep[][and references therein]{mai17}, whose spectra  match several of the DIBs.
  
 In this paper we concentrate on the unidentified, broad absorption in the $3.2-3.3~\mu$m spectral interval. 
 Considerable analysis of the numerous unidentified emission and absorption bands at wavelengths $\geq$3.3 $\mu$m, which are prominent toward a wide range of galactic and extragalactic sources, has been done, as discussed in some of the review papers referred to above. However, there has been little analysis of the accompanying absorption present in the $3.2-3.3$~$\mu$m interval.   A broad absorption feature in this interval was discovered toward the young stellar object (YSO) Mon R2 IRS3 \citep{sel94, sel95} and subsequently was detected toward additional YSOs in the Galactic plane via both ground-based \citep{bro96, bro99} and space-based \citep{bre01} spectroscopy. Hereafter, we refer to this absorption seen toward YSOs as the 3.25 $\mu$m  band, to distinguish it from the absorption analyzed in our paper, which may arise in a different environment and may have a different spectral profile. The intrinsic weakness of the absorptions observed in the $3.2-3.3~\mu$m interval and  other observational challenges described below have made it difficult to characterize the absorption in much detail. 

 The $3.2-3.3~\mu$m spectral interval encompasses a large portion of the aromatic CH stretching vibration, hereafter referred to as  CH(S), characteristic of polycyclic aromatic hydrocarbons (PAHs).  PAHs are generally considered to be the most abundant of the large interstellar molecules, an interpretation that is commonly referred to as the PAH hypothesis \citep{all89,tie08}. If that hypothesis is correct and if one can accurately determine the underlying in-terstellar molecular absorption spectrum, it may be possible to constrain some properties of the interstellar PAHs, including size, symmetry, physical state, and charge. 

Most of the sources toward which the 3.25~$\mu$m band has been observed to date are YSOs, i.e., stars that are situated within the dense molecular clouds out of which they formed. Some  of the papers reporting the detections in those objects have tentatively attributed the absorption to free aromatic molecules in cold interstellar molecular gas \citep[][and references therein]{bro99}. Note, however, that diffuse interstellar gas may exist exterior to the dense clouds in which the YSOs are located and could contribute to the observed absorption. 

The 3.25~$\mu$m absorptions detected from the ground with low signal-to-noise ratios (S/Ns) by \citet{sel95} and \citet{bro96,bro99} can be approximately fit by single Gaussians with central wavelengths close to 3.25~$\mu$m and full widths at half maximum (FWHMs) of $\sim$0.07~$\mu$m. In contrast, the 3.25~$\mu$m absorptions observed toward YSOs by the International Space Observatory (ISO) and analyzed by \citet{bre01} appear to have a wider range of central wavelengths, widths, and absorption profiles, and most of them do not appear amenable to fitting by single Gaussians. Due to the low S/Ns of the absorptions in all of these data it is unclear at present to what extent these source-to-source variations toward YSOs are real.
 
  More recently, a broad absorption band extending from 3.2 to 3.3~$\mu$m has been clearly detected on two sightlines toward the Galactic center (GC) that are known to contain large column densities of diffuse gas \citep{whi97}. The spectra on GC sightlines are toward 2MASS $J$17470898-2829561 (2M1747), observed by \citet{geb21} at the Gemini North Telescope, and toward the Quintuplet Cluster (QC), observed by \citet{chi13} at the United Kingdom Infrared Telescope. The latter authors published a multi-source-averaged spectrum of the QC. Tentative detections of absorption in this wavelength interval toward three GC sources were previously reported by \citet{pen94}. The absorption also appears to be present  in the spectrum obtained by the Short Wavelength Spectrometer of the Infrared Space Observatory (ISO) of the diffuse cloud in front of the widely investigated star Cygnus OB2-12 \citep{hen20}, whose sightline is much shorter and lies in a different direction than that toward the GC. In addition, although not reported as detected, the absorption may be present at low S/N in the spectrum of the diffuse cloud on the sightline to the star S7 located in the IRAS 18511+0146 cluster \citep{god12}.  
  
Rather than peaking near 3.25~$\mu$m as is the case for the spectra of the feature toward YSOs obtained by \citet{sel94, sel95} and \citet{bro96, bro99}, the peak optical depth, $\tau_{\rm p}$, of the band toward the above GC sources appears to occur near 3.28~$\mu$m.  Sightlines toward the GC are known to pass through both diffuse clouds and dense clouds, the latter associated with one or more of the intervening three spiral arms \citep{whi97}. The GC's Central Molecular Zone is also known to contain large amounts of dense and diffuse gas \citep[][and references therein]{mor96,oka19}. The depths of the $3.2-3.3~\mu$m absorption and the adjacent and well-studied 3.4~$\mu$m  absorption band toward the QC and toward 2M1747 \citep{chi13,geb21} are nearly identical; however, the visible extinction toward the former is $\sim$30 mag \citep{fig99}, whereas the extinction toward the latter is $\ge$100 mag \citep{geb21}. Because most of the extinction to 2M1747 is due to dense cloud material, the near equality of these two absorptions on the two GC sightlines indicates that they arise mainly in diffuse clouds.

 Accurate spectroscopy of  the $3.2-3.3~\mu$m spectral region from the ground is observationally challenging.  Because of the intrinsic weakness of the interstellar absorption, one must seek bright yet distant IR sources behind large column densities of gas. More importantly, strong atmospheric absorption lines of H$_2$O and CH$_{4}$ cover much of this spectral interval.  CH$_{4}$ is uniformly mixed in the atmosphere, so that its observed column density depends only on the zenith angle of the observation.  In contrast, the H$_{2}$O mixing ratio can  vary spatially and temporally (i.e., due to weather).  These considerations make accurate extraction of the interstellar absorption spectrum from terrestrial-based observations problematic, even if the airmasses of the target object and the calibration star are nearly identical. The final spectrum, if reduced using standard procedures, can easily exhibit residual H$_{2}$O and CH$_{4}$ features corresponding to under- or overcorrection. Because most of the telluric H$_{2}$O and CH$_{4}$ lines in the $3.2-3.3~\mu$m interval have substantial optical depths, even small fractional changes in the atmospheric H$_{2}$O and CH$_{4}$ column densities can produce large residual features.  The ISO spectrum of the diffuse cloud in front of Cyg OB2-12 \citep{hen20} does not suffer from this atmospheric absorption problem.  However, in the $3.2-3.3~\mu$m region the ISO spectrum has a low S/N and requires substantial smoothing  in order to obtain a meaningful profile of the absorption for use in further spectral analysis.
  
 This paper is organized as follows.  First we describe and apply our method of correcting the $3.2-3.3~\mu$m portion of the published spectrum of 2M1747 for the different airmasses at which it and its telluric standard were observed. Then, using a superposition of Gaussians, we obtained a good fit to the corrected 2M1747 spectrum, and also to the  $3.2-3.3~\mu$m segments of the QC and Cyg OB2-12 spectra. In the discussion section we compare our fits to previous analyses of the $3.2-3.3~\mu$m absorptions. We then discuss and justify our identification of different classes of PAHs with individual Gaussian components of the multi-Gaussian fits. We also estimate PAH sizes and abundances, and apply a PAH cluster (i.e., PAH dimers, trimers, ...) formation model to predict the relative abundances of gas-phase monomer PAHs and condensed-phase PAH molecular clusters. The relative abundances of PAHs in these phases bear on the smoothness of the retrieved spectrum. Following the summary are two appendices, one that derives the correction factor for the airmass residual optical depth, and the other that derives the depletion of gas-phase, monomer PAHs and the formation of PAH molecular clusters in the envelopes of post asymptotic giant branch (post-AGB) stars.

\begin{figure}
\begin{center}
\includegraphics[width=0.45\textwidth, angle=-0]{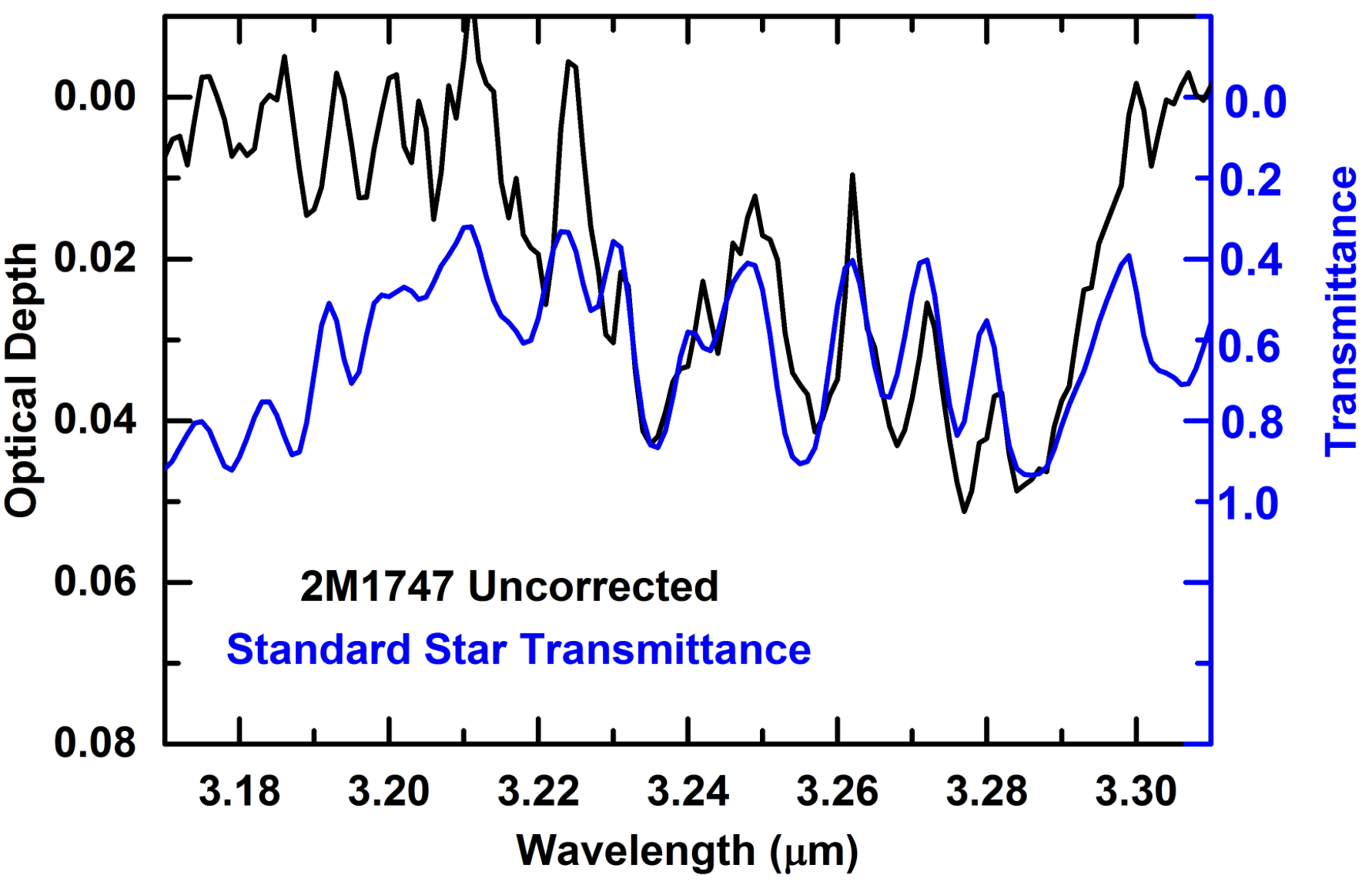}
\end{center}
\caption{Comparison of originally reduced and published (Uncorrected) optical depth spectrum of 2M1747 in the $3.17-3.31~\mu$m interval \citep[black line, from][]{geb21} to the scaled transmittance (right hand scale; transmittance increases downward) of earth's atmosphere (blue line) in the same interval, derived from the telluric standard for 2M1747, HIP 89622, observed at 0.06 higher airmass. The spectral resolution was 0.0036~$\mu$m (R = 900).  Note the strong correlation between several local minima in optical depth and minima in transmittance,}
\end{figure}

\section{DATA  REDUCTION AND SPECTRAL FITTING}

\subsection {2MASS J7470898-2829561 (2M1747)}

    The starting point for the analysis of the $3.2-3.3~\mu$m spectrum of  2M1747 was the initially-reduced optical depth ($\tau$) spectrum shown in Figure 1, obtained from  Figure 9 of \citet{geb21}. The resolving power (R = $\lambda/\Delta\lambda$) of this spectrum is  900. Their initial data reduction steps included the usual practice of differencing nodded pairs of spectral images to remove atmospheric emission. Flux calibration and approximate correction for the atmospheric transmittance ($t$) in the extracted spectrum of 2M1747 were achieved by division of this extracted spectrum by the similarly extracted spectrum of the standard star, HIP 89622. 2M1747 was observed at a mean airmass of 1.52; the standard star at a mean airmass of 1.58.  At first glance, the spectral structure in the 2M1747 spectrum in Figure 1 might be attributed to noise.  However, as also seen in Figure 1, there is a high degree of correlation between the 2M1747 spectral features and the standard star telluric transmittance also plotted in the figure; local minima in transmittance correspond to local minima in $\tau$.  (Due to the heavily scaled-down transmittance in Figure 1 the amplitudes of many of the features in the two  spectra appear comparable.) During the period of observations of 2M1747 and the standard star, the Maunakea water vapor column at zenith was steady at 1.5 $\pm$ 0.1 mm.
                
\begin{figure}
\begin{center}
\includegraphics[width=0.45\textwidth, angle=-0]{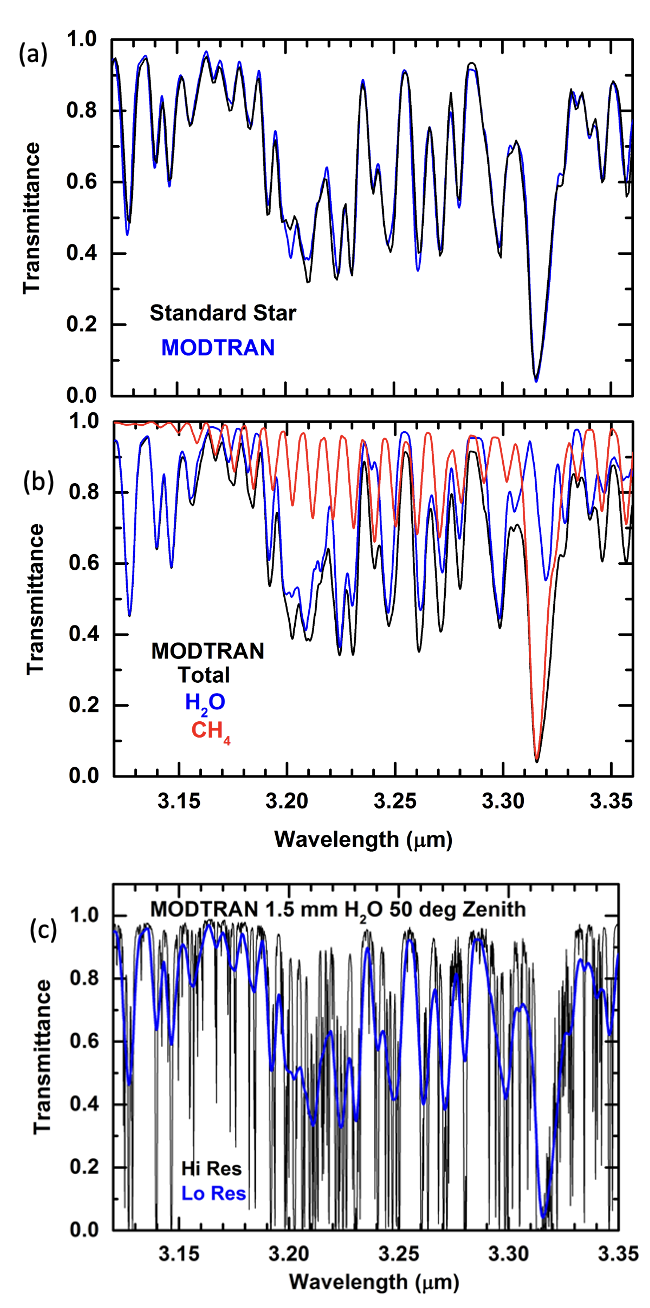}
\end{center}
\caption{(a) Comparison of the model spectrum computed using MODTRAN \citep{ber17} to the transmittance spectrum of the standard star used in the 2M1747 data reduction, in the interval $3.12-3.36~\mu$m. The MODTRAN calculations were based on a triangular slit function with R = 900 (FWHM = 0.0036~$\mu$m at 3.25~$\mu$m), and a  zenith angle of 50 deg.  Columns of atmospheric H$_{2}$O and CH$_{4}$ were adjusted to fit the standard star spectrum. The model H$_{2}$O column was determined to be 1.5~mm, consistent with the measured conditions on Maunakea during the observations. (b) Individual contributions of atmospheric H$_{2}$O and CH$_{4}$ in the MODTRAN transmittance spectrum, for R = 900.  (c) Comparison of modeled R = 30,000 and R = 900 transmittance spectra. The high-resolution calculation shows that the spectrum is mainly composed of very opaque, narrow lines (i. e., transmittances $\lesssim$0.1) separated by narrow intervals of high transmittance, typically $\sim0.95$ when observed at R = 900.}
\end{figure}
    
 The standard star telluric transmittance used for the 2M1747 data reduction is compared to a model transmittance calculation in Figure 2a.  The model-based calculation was used to define the stellar continuum and to fine tune the wavelength calibration for the standard star transmittance spectrum, which was then applied to the 2M1747 optical depth spectrum.  The standard star transmittance was tied to intervals of maximum transmittance in the model calculation, which were $\sim$0.95 in the $3.12-3.36~\mu$m interval shown in Figure 2a.  Given the similarity of the two transmittance results, either could have been used in the analysis described below.  However, there are a few minor differences; the observed transmittance was slightly preferred for the analysis.  
  
The difficulty of retrieving accurate interstellar absorption spectra from a terrestrial observation in the $3.2-3.3~\mu$m spectral region stems from the challenge of accurately removing the spectrally structured transmittance spectrum caused by a multitude of strong H$_{2}$O and CH$_{4}$ absorption lines, whose individual and combined contributions are shown in Figure 2b.  The optical depths of the deepest features in this interval are -ln(0.38) $\sim$ 1 as observed at R = 900.  This means that even airmass differences as small as 1\%\, between the standard star and source object sightlines, can result in residual optical depths in the ratioed spectrum of order 0.01, already comparable to the ``apparently" random noise in the 2M1747 optical depth spectrum in Figure 1.

\begin{figure}
\begin{center}
\includegraphics[width=0.45\textwidth, angle=-0]{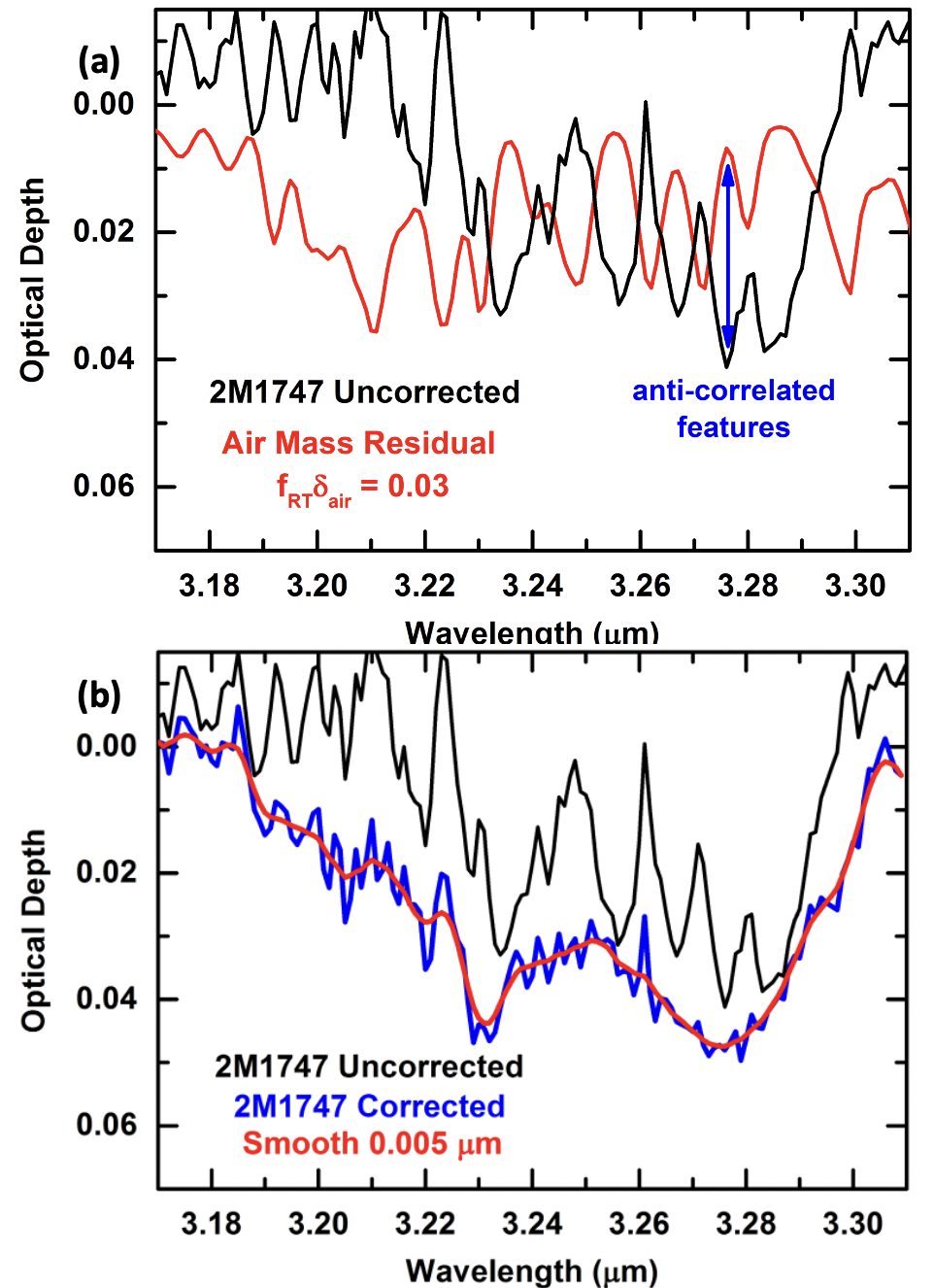}
\end{center}
\caption{(a) Comparison of airmass optical depth residual spectrum (red) and uncorrected 2M1747 optical depth spectrum (black) from Figure 1. As discussed in the text, we shifted the uncorrected spectrum vertically by 0.01 relative the the spectrum in Fig. 1. to approximately account for residual telluric H$_{2}$O emission features, which must be present weakly across the entire spectrum including the short wavelength portion of the spectrum.  The residual optical depth corresponds to a 0.06 airmass difference (see text for details).  The vertical blue line indicates one of several instances of anti-correlation of optical depth residual spectral features  with emission-like features in the uncorrected 2M1747 spectrum.  (b) Airmass residual uncorrected (black) and corrected 2M1747 (blue) spectra; the latter is the sum of the two spectra the upper panel. The noise in the corrected spectrum (judged by point-to-point fluctuations) is significantly larger at the shorter wavelengths due to generally lower atmospheric transmittance (requiring a larger correction) there. The red spectrum is the corrected spectrum with a five-point ($0.005~\mu$m) triangular smoothing (resulting in an overall resolution of 0.0062~$\mu$m), which was used in the spectral fitting process.}
\end{figure}

 As detailed in Appendix A, the correction of an observed optical depth spectrum for the fractional airmass difference, $\delta_{air}$ (i.e., the ratio of the airmass difference and airmass of the standard star), is expressed in terms of the residual optical depth, $\Delta\tau_{air}$, as
\begin{equation}
\Delta\tau_{air} =  f_{RT} \delta_{air} \tau_{line} + \delta_{air} \tau_{cont},
\end{equation}

\noindent where  0.5 $\le$ $f_{RT}$ $\le$ 1.0 is a radiative transfer (RT) factor (see Appendix A),  and $\tau_{line}$ and $\tau_{cont}$ are the molecular line and continuum optical depths obtained from the standard star transmittance.  The residual airmass can be positive or negative, resulting in either dips or bumps in the ratioed spectrum. The value of  $f_{RT}$  depends on the distribution of the optical depths of the telluric lines in each spectral resolution element.  For lines that are either close to or in the strong line limit, i.e. $\tau_{\rm{p}} \gg 1$ for pressure-broadened (Lorentz) lines,  $f_{RT}$ $\approx$  0.5. The  widths of the H$_{2}$O and CH$_{4}$ lines in Figure 2c are on the order of 0.00003 $\mu$m (0.03 cm$^{-1}$), which is far narrower than the observed spectral resolution of 0.0036 $\mu$m (3.4 cm$^{-1}$).  As can be seen in Figure 2c, many of the telluric lines in the $3.2-3.3~\mu$m spectral interval are at or near the strong-line limit.   However, the intervals with the strong lines also contain weaker lines and overlapping line wings, which are in the weak line limit.  Consequently, the values of $f_{RT}$ for these intervals fall between the two limiting values of 0.5 and 1.0.  In Appendix B, we find that treating $f_{RT}$ as a fitting parameter results in an average value for the  $3.2-3.3~\mu$m interval of  0.79.      

The airmass residual optical depth spectrum for the 2M1747 measurements, with $f_{RT}$ = 0.79 and  $\delta_{air} = +0.038$ (i.e.,   $f_{RT}\delta_{air}$ = 0.030) is compared to the uncorrected 2M1747 optical depth spectrum in Figure 3a. In this figure we have shifted the optical depth scale of the uncorrected spectrum by 0.01 with respect to the ratioed spectrum in Figure 1. This is to approximately account for the presence of residual emission in the ratioed spectrum due to incomplete cancellation of telluric lines. The residual emission exists across the entire spectral region being investigated, but is only clearly visible in the $3.21-3.28~\mu$m interval in Figure 1 as correlations between the very strongest telluric absorption features and emission features in the optical depth spectrum. In particular, due to the airmass mismatch there must be weak residual emission (i.e., negative optical depth) in the optical depth spectrum in the $3.17-3.19~\mu$m interval, which lies just to shorter wavelengths from the $3.2-3.3~\mu$m interstellar band, because moderately strong lines of both H$_{2}$O and CH$_{4}$ are present there (see Figure 2). 

As can be seen in Figure 3a, the residuals have approximately the same strengths as the corresponding features in the uncorrected spectrum, but are anticorrelated with them. This is as expected since the standard star was observed at a higher airmass than was 2M1747.  Addition of the two spectra result in the corrected 2M1747 optical depth spectrum (blue curve) shown in Figure 3b.    This spectrum appears to consist of  broad absorptions overlaid with measurement noise.  A mild, five-point (0.005~$\mu$m) running smooth of the corrected spectrum produced the Smooth spectrum in Figure 3b (red curve), which was adopted for the spectral fitting described next.  Note that  as expected, there is little or no interstellar absorption at $3.17-3.19~\mu$m, which indicates that our offset of 0.01 in the previously published optical depth spectrum is appropriate.

Inspection of the smoothed 2M1747 spectrum in Figure 3b suggests that three or four Gaussians are needed to fit it.  We performed a visual  fit with the simplification that each Gaussian could be sequentially ``peeled off" and fit separately.  For example, Figure 4 shows that the long wavelength half of the longest wavelength Gaussian is only slightly overlapped by the adjacent Gaussian.  From this half-Gaussian, one can obtain a good initial estimate of the height, width, and position of the longest wavelength fitted Gaussian.  With that fit in hand, the parameters for the adjacent Gaussian could then be determined.  This procedure was repeated until the entire spectral interval of interest was fitted.  We made a final fitting pass to more accurately account for the spectral overlap between the Gaussians.  The best four-Gaussian fit is shown in Figure 4. For the other two sightlines discussed below, the initial values of the fitting parameters were based on those for the 2M1747 fit.  The parameters of the fits for all three sightlines, and estimates of their uncertainties, are presented in Table 1 and their identifications are discussed in Section 3.2.

\begin{table*}
\begin{center}
\caption{Gaussian Fit Parameters$^{a}$}
\begin{tabular}{ccccccccc}
\hline\hline
2M1747 & 2M1747 & 2M1747 &  Quintuplet & Quintuplet & Quintuplet & Cyg OB2-12 & Cyg OB2-12 & Cyg OB2-12 \\
Band Center & FWHM & $\tau_{\rm p}$  &Band Center & FWHM & $\tau_{\rm p}$ & Band Center & FWHM & $\tau_{\rm p}$ \\   
($\mu$m / cm$^{-1}$) & ($\mu$m / cm$^{-1}$) & & ($\mu$m / cm$^{-1}$) & ($\mu$m / cm$^{-1}$) & & ($\mu$m / cm$^{-1}$) & ($\mu$m / cm$^{-1}$) & \\
\hline
3.204 & 0.022 & 0.0187 & 3.205 & 0.020 &  0.0118 & 3.209 & 0.0186 & 0.0045 \\
3112.1 & 21.4 & & 3120.6 & 19.5 & & 3116.4 & 18.1 & \\
\hline
3.2308 & 0.024 & 0.0346 & 3.232 & 0.026 & 0.0268 & 3.2384 & 0.0224 & 0.0104\\
3095.2 & 23.0 & & 3094.0 & 24.9 & & 3087.9 & 21.4 & \\
\hline
3.2573 & 0.0314 & 0.0269 & 3.265 & 0.036 & 0.0281 & 3.264 & 0.036 & 0.0070 \\
3070.0 & 29.6 & & 3062.8 & 33.8 & & 3062.8 & 33.8 & \\
\hline
3.281 & 0.030 & 0.0398 & 3.282 & 0.036 & 0.0476 &  3.282 & 0.037 & 0.0105 \\
3047.9 & 27.9 & & 3046.9 & 33.4 & & 3046.9 & 34.3 & \\
\hline\hline
\end{tabular}
\end{center}
$^{a}$ Estimated uncertainties for central wavelengths, FWHMs, and $\tau_{\rm p}$ are 0.020~$\mu$m, 0.020~$\mu$m, and 5\%., respectively \\
\end{table*}

\begin{figure}
\begin{center}
\includegraphics[width=0.45\textwidth, angle=-0]{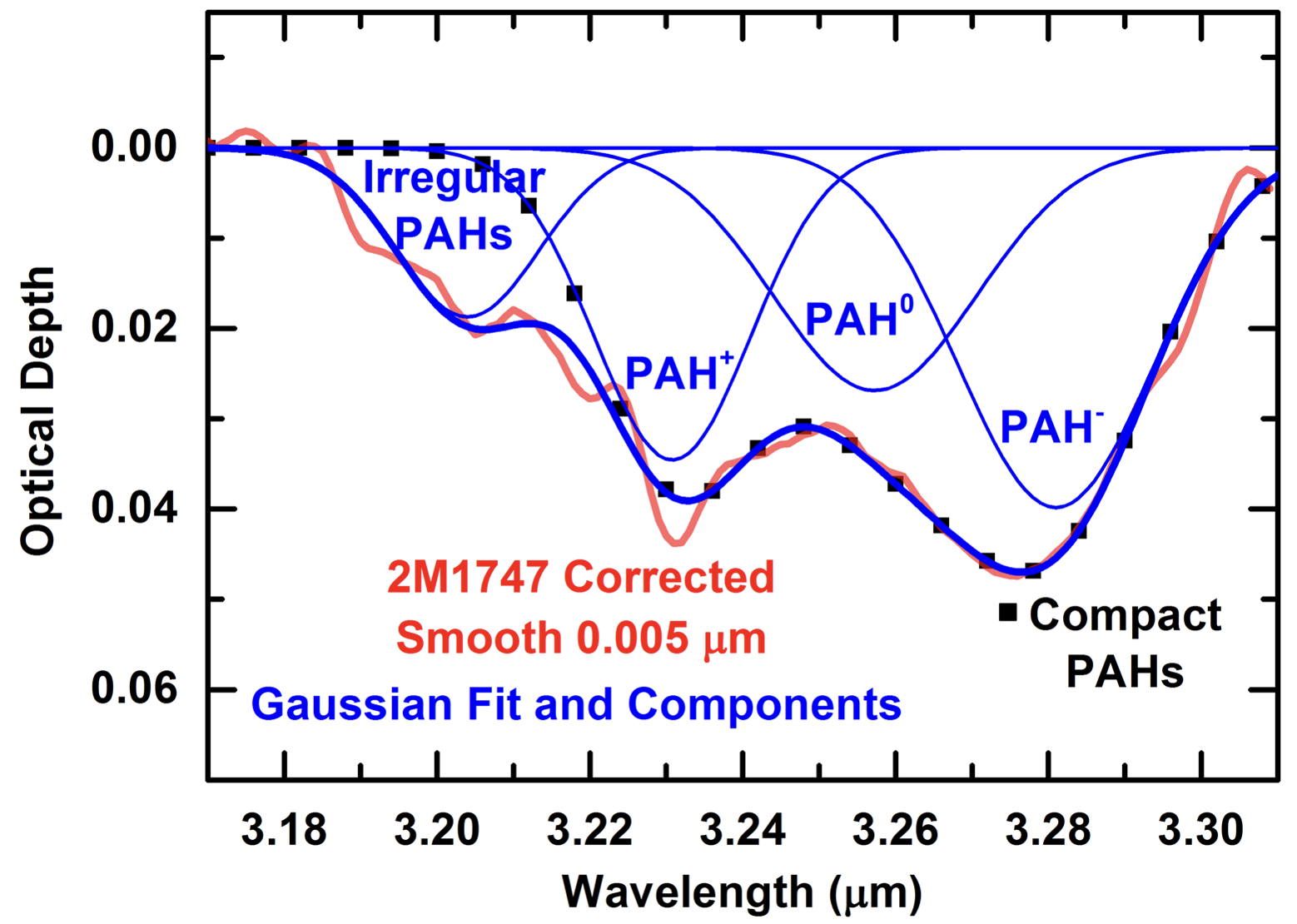}
\end{center}
\caption{Comparison of the corrected, smoothed 2M1747 spectrum (red curve) and four-Gaussian fit (thick blue curve).  Also displayed are the individual component Gaussians in the fit (thin blue lines). The component Gaussians can be associated with specific PAH symmetry types, compact and irregular, and charge states, cationic, neutral, and anionic as labeled (see Section 3.2 for details), The sum of the compact PAHs is denoted by the thin black curve passing through the filled squares. See Table 1 for fit parameters.}
\end{figure}

\subsection{Quintuplet Cluster}

 The analysis of the $3.2-3.3~\mu$m spectrum of QC utilized the multi-source-averaged spectrum from Figure 2 in \citet{chi13}, observed at R =  1,400 and shown in Figure 5.   The spectral profiles of the absorption in the individual sources used to produce this average spectrum have significant differences (see their Figure 1) and are noisy, which may attest to the inherent difficulties in ratioing ground-based  spectra in this wavelength interval, as discussed previously. However, there is little or no correlation between the strongest narrow spectral features at 3.215~$\mu$m, 3.237~$\mu$m, and 3.252~$\mu$m in the QC spectrum with strong telluric absorption lines (see Figure 1). \citet{chi13} utilized a number of telluric standards in their calibration, whose differences in airmass from the QC sources were very small, $0.01-0.02$. Based on the above considerations, unlike for 2M1747 we did not attempt to correct the published QC spectrum for telluric residuals.  
 
 \begin{figure}
 \begin{center}
\includegraphics[width=0.45\textwidth, angle=-0]{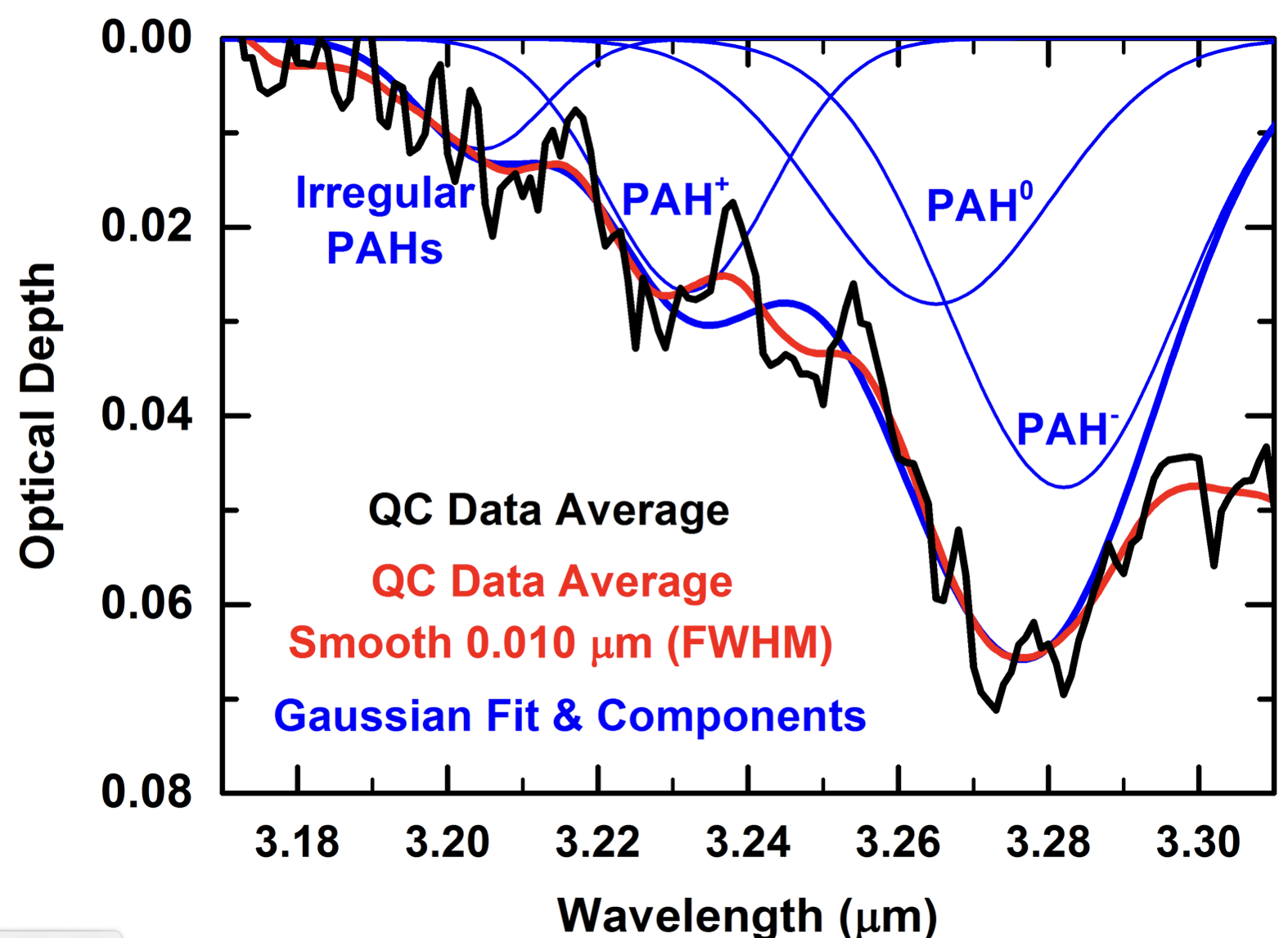}
\end{center}
\caption{Comparison of the smoothed optical depth spectrum of the Quintuplet Cluster (red curve) and a four-Gaussian fit (thick blue curve) made using the same approach as for 2M1747 (see text and Figure 4 caption). The original unsmoothed optical depth spectrum from \citet{chi13} is shown in black..}
\end{figure}

  A significant difference between the shapes of the optical depth spectra of 2M1747 and QC is their behaviors near 3.30~$\mu$m, where the 2M1747 spectrum shows a much smaller optical depth than the QC spectrum. This appears to be due to different techniques used to define the continuum. For 2M1747 \citet{geb21} estimated a continuum level at 3.34~$\mu$m, which they assumed was depressed only by the wing of the 3~$\mu$m ice band, for which they corrected, and pinned their optical spectrum to that wavelength along with other wavelengths far from $3.2-3.3~\mu$m. \citet{chi13} used as their continuum a third order polynomial pinned to the narrow short and long wavelength segments at near 3.20~$\mu$m and 3.65~$\mu$m; as can be seen in their Figure 1; doing so results in significant optical depth in the $3.30-3.34~\mu$m interval, which they ascribe to a blending of the long wavelength wing of a single broad Gaussian centered at 3.28~$\mu$m fitting the $3.2-3.3~\mu$m absorption and the short wavelength wing of the shortest wavelength aliphatic component of the 3.4~$\mu$m absorption band. 

 Because the sightlines to 2M1747 and QC are very close and therefore are likely to pass through diffuse interstellar gas with similar properties, we began the fitting procedure using the Gaussian wavelengths and widths derived for 2M1747. To assist in the iterations we applied a 0.010~$\mu$m smoothing to the published optical depth spectrum. Due to the differences in the derivation of absorption optical depth for 2M1747 and QC discussed above  we  did not attempt to fit the published spectrum longward of 3.29~$\mu$m. The four-Gaussian fit to the smoothed QC spectrum is shown in Figure 5.
 
 \begin{figure}
 \begin{center}
\includegraphics[width=0.45\textwidth, angle=-0]{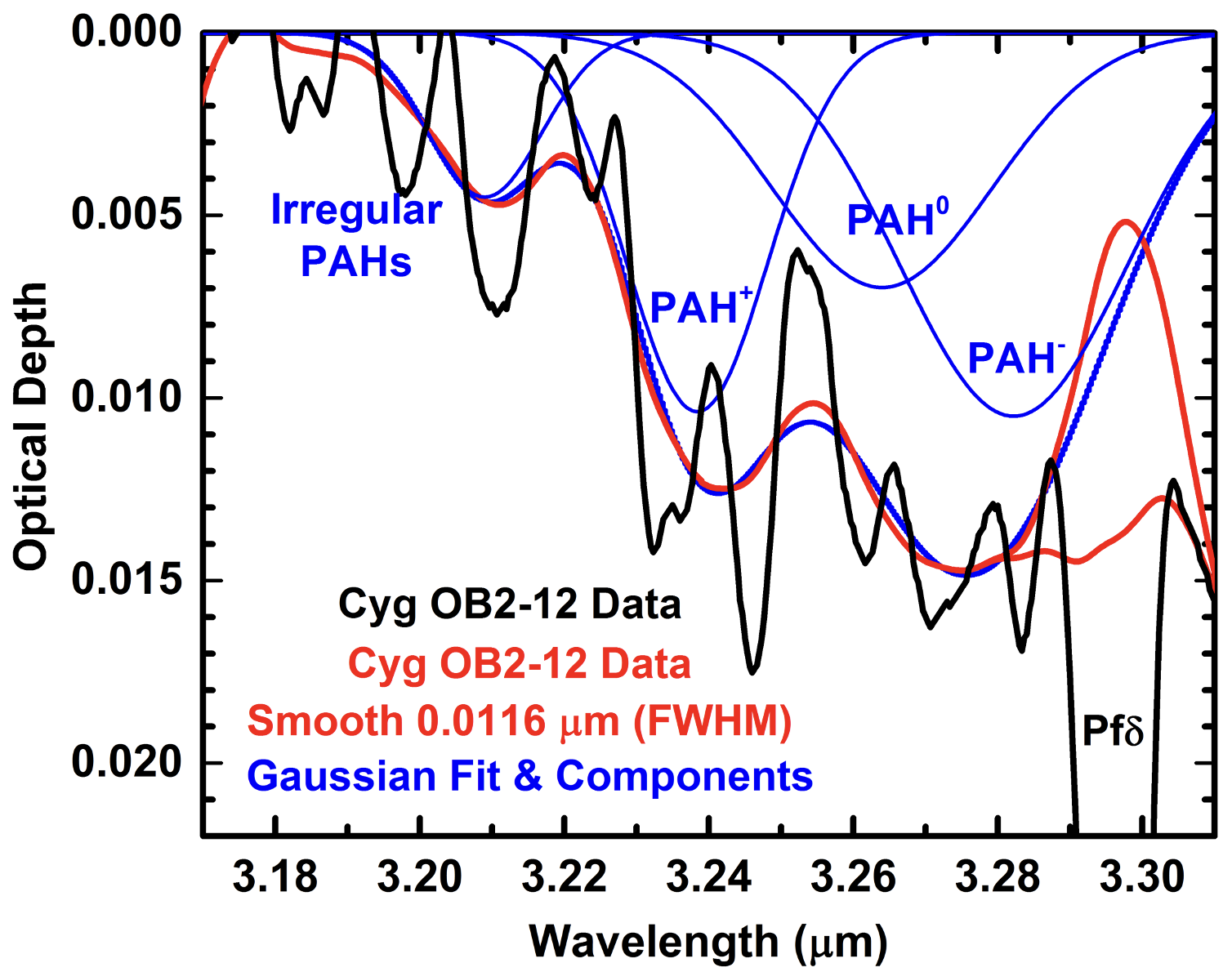}
\end{center}
\caption{Comparison of the  smoothed optical depth spectrum of the Cyg OB2-12 sightline (red curves) and a four-Gaussian fit (thick blue curve), made using the same approach as for 2M1747 and QC  (see text). The original optical depth spectrum from \citet{hen20}, shown in black, has a resolution (FWHM) of 0.0052~$\mu$m. The additional 0.012~$\mu$m smoothing that produced the red curve resulted in a resolution for it of 0.014~$\mu$m. The relative contributions of the strong 3.297~$\mu$m  Pf~$\delta$  and the interstellar opacity near that wavelength are not well known, especially in the wings of the line.  Gaussian fits for two limiting cases are shown for the interstellar opacity longward of  3.285~$\mu$m, one assuming no interstellar opacity and one assuming an interstellar opacity matching the observed value at 3.285~$\mu$m. The effects of these baseline choices correspond to the different curves (red lines) beyond 3.285~$\mu$m.  This uncertainty did not impact the Gaussian fit, since only the wavelength region at wavelengths less than 3.285~$\mu$m was used in the fitting procedure.}
\end{figure}

\subsection{Cygnus OB2-12}

Our analysis of the $3.2-3.3~\mu$m absorption toward  Cyg OB2-12 observed by ISO is shown in Figure 6. We used the optical depth spectrum of Cyg OB2-12 shown by \citet{hen20} in their Figure 7 at a resolution of 0.0052~$\mu$m. Because of the very low S/N of the interstellar absorption, we smoothed the spectrum, by 0.012~$\mu$m, in order to more easily facilitate the Gaussian fitting. The overall resolution of the smoothed spectrum is 0.014~$\mu$m  As detailed in Figure 6, we also removed, by Gaussian fitting, several limiting cases of the strong Pf~$\delta$ line at 3.297~$\mu$m from the optical depth spectrum of Cygnus OB2-12. Because of the uncertainty in the relative contributions of Pf~$\delta$ and the interstellar opacity, we did not attempt to fit the spectrum at wavelengths longer than 3.285~$\mu$m. The smoothed optical depth spectrum shortward of 3.285~$\mu$m was fitted, primarily through adjustment of the peak optical depths of the four Gaussians used for the 2M1747 fit. The best-fit parameters for Cyg OB2-12 are given in Table 1.

\begin{figure}
\begin{center}
\includegraphics[width=0.45\textwidth, angle=-0]{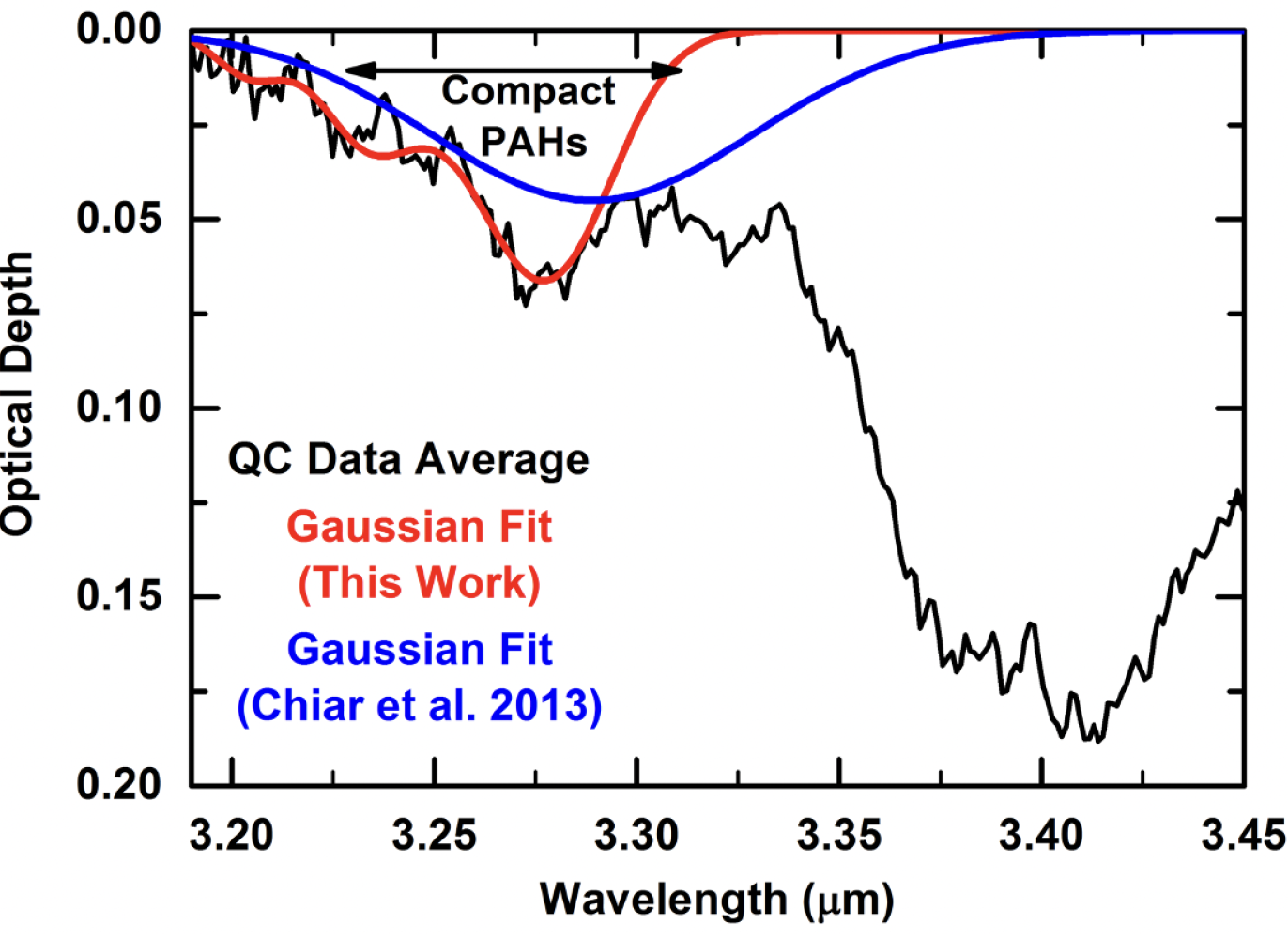}
\end{center}
\caption{Comparison of Gaussian fits to the QC optical depth spectrum (black line) from this work (red line) and \citet[][blue line]{chi13},  Also shown by the double-arrow line is the spectral interval for compact PAHs, as determined from the NASA PAH Database.}
\end{figure}

\section{DISCUSSION}

\subsection{Previous Analyses of the Absorption Band}

  Spectral fits of the $3.2-3.3~\mu$m absorption  band in two of the sources considered here, QC and Cyg OB2-12 have been performed previously \citep{chi13,hen20}.  In both cases, a single, broad Gaussian was used. The current multi-Gaussian fit to QC and the previous single-Gaussian fit are shown in Figure 7.   The single Gaussian is quite wide, with FWHM $\approx$ 0.09~$\mu$m ($\approx$ 80 cm$^{-1}$) and its long wavelength wing  extends well beyond the long wavelength limit for which aromatic CH(S) features are centered, $3.22-3.31~\mu$m (see Section 3.2), as indicated by the double arrow line in Figure 7.   The same comment applies to the single Gaussian fit to the $3.2-3.3~\mu$m band for Cyg OB2-12 \citep{hen20}, which was based on the \citet{chi13} fit.  Both the QC and Cyg  OB2-12 fits have significant absorption in the $3.30-3.34~\mu$m interval, whereas 2M1747 fit has essentially no absorption in the vicinity of 3.30~$\mu$m.  We suggest that absorption in this interval, if present, is not due to aromatic CH(S) but possibly to super-hydrogenated PAHs \citep{yan20} and/or olefins, as inferred from laboratory measurements of carbonaceous films \citep{gie96}.  As discussed earlier for QC, this absorption may be in part, or completely, due to the method used for the continuum removal.  Additionally, the single-Gaussian fit does not account for the relatively narrow absorption peak at 3.28~$\mu$m. \citet{chi13} assigned this feature to olefins.  Our interpretation (Section 3.2) is that it is due to the high relative abundance of the PAHs producing the longest wavelength component Gaussian.
  
 \citet{hen20} noted that the widths of the single Gaussians fitted to the QC and Cyg OB2-12 spectra were inconsistent with the much narrower width, typically 0.041~$\mu$m (38 cm$^{-1}$), observed for the 3.3 $\mu$m emission feature \citep{tok21}.  Our analysis addresses this inconsistency, since we find that the $3.2-3.3~\mu$m absorption band can be attributed to four Gaussians, with relatively narrow widths of $0.02-0.03~\mu$m (20-30 cm$^{-1}$).  As noted above, in our interpretation the IR absorption is attributed to a superposition of charged and neutral compact PAH clusters, where the broadening is a consequence of PAH-PAH potential interactions in the low-temperature condensed, molecular cluster phase \citep{ros15}.   Our multi-Gaussian fit is consistent with the widely held view that the 3.3~$\mu$m emission feature is produced by the overlap of emission features from gas-phase PAH cation and neutral monomers, where the broadening and band shifts are a consequence of the high emission temperatures \citep{job94,job95,tok21}.
    
 \subsection{Identification of the Gaussian Fit Components with Classes of PAHs}
 
 As indicated in Figures 4-6, we attribute the three longest wavelength Gaussians primarily to compact PAHs in the three charge states: cations, neutrals, and anions.  According to the NASA PAH Database \citep{boe14,mat20} compact PAHs in these charge states are predominantly confined to the $3084-3034$ cm$^{-1}$ ($3.24-3.30~\mu$m) spectral interval.  There are only a few cm$^{-1}$ differences in the band locations for compact PAHs in the N$_{\rm{C}}$ $\sim$ 50-100 size range (where N$_{\rm{C}}$ is the number of carbon atoms) attributed to interstellar PAHs \citep{tok21,kni21}.  However, as discussed below, the three charge states are confined to different spectral sub-intervals.  Extending the upper and lower spectral boundaries by 15 cm$^{-1}$ to account for the uncertainty in Density Functional Theory (DFT) calculations \citet{bau18} confines the compact PAHs to the spectral range 3099-3019 cm$^{-1}$ ($3.227-3.315~\mu$m).  This spectral interval is consistent with the central wavelengths of the three longer wavelength Gaussians in Figures 4-6 and listed in Table 1.

Because of their high degrees of symmetry, compact PAHs only contain solo (isolated) CH bonds, and duo (two adjacent) CH bonds with no neighboring CH bonds.  As a specific example of the locations of CH(S) fundamental bands of compact PAHs, we consider the proposed parent carrier, circumovalene (C$_{66}$H$_{20}$), discussed in the following subsection. In the NASA PAH Database it has cation, neutral, and anion bands at 3084, 3071, and 3055 cm$^{-1}$ (3.243, 3.257, and 3.273 $\mu$m), respectively. In comparison, the three longer wavelength Gaussians in the fit to 2M1747 are at 3096, 3071, and 3048 cm$^{-1}$ (3.230, 3.257, and 3.281 $\mu$m). The two sets of central wavelengths are consistent with each other to within the  $\pm$ 15 cm$^{-1}$ ($\pm$ 0.016~$\mu$m ) DFT uncertainty \citep{bau18}.  

Irregular PAHs can contain solo, duo, trio and quartet CH bonds. The solo and duo CH(S) bonds coincide with their charged and neutral compact PAH counterparts. As a specific example of the locations of the trio and quartet bonds, we consider the irregular PAH C$_{66}$H$_{24}$, which after broadening by a Gaussian with FWHM  20 cm$^{-1}$, has features centered at 3133, 3125, and 3110 cm$^{-1}$ (3.192, 3.200, and 3.215 $\mu$m) for the cation, neutral, and anion, respectively. To within DFT uncertainties, these irregular bands are all viable contributors to the shortest wavelength Gaussian toward 2M1747, QC, and Cyg OB2-12 in Table 1 at 3.210, 3.205, and 3.210 $\mu$m, respectively. However, the separation of the above DFT wavelengths is greater than the observation-derived FWHMs of the shortest wavelength Gaussian on all three sightlines, which suggests that fitting this feature with a single Gaussian is unrealistic. Due to the low S/N on this weakest portion of the $3.2-3.3~\mu$m absorption, we decided not to attempt a more complex Gaussian fit to the putative irregular absorption feature.  
  
 One would expect that the CH(S) spectrum of a PAH on or in a grain (e.g., a cluster) would be broadened and shifted relative to its  gas phase spectrum.  Laboratory IR absorption spectroscopy of PAHs deposited on a substrate at 5 K provide  a direct measure of the shifts and widths for a number of small neutral PAHs \citep{ros15}. They found that a 5 K, thin film of pure pyrene, C$_{16}$H$_{10}$, had a width of $\sim$20 cm$^{-1}$ and a shift of $\sim$$-6$ cm$^{-1}$ relative to an isolated pyrene molecule in Ar at 5 K.  Shifts and widths of such magnitudes for larger, clustered PAHs would be consistent with the observed broad and smooth IR absorption spectra.  
 
The similarities of the wavelengths and widths of the three Gaussians corresponding to compact PAHs on the three sightlines analyzed here supports the notion that each of the sightlines contains similar sets of PAHs.  The source-to-source variations in the relative strengths of the Gaussian components  makes sense if each feature corresponds to a particular charge state. The fractional abundance of a charge state is sensitive to the properties of its local environment, such as density, temperature, and ultraviolet (UV) flux \citep{wei01}.  \citet{li01} estimated that for carbon grains distributed along an average sightline and in the size range of the PAH clusters, $\sim$330 carbon atoms, approximately 50\% of the grains would carry a charge.  This fraction of charged PAH clusters is similar to that determined from the Gaussian fits to the IR absorption spectra (see Section 3.3.2 below for details).  The distribution of environments will be unique for each sightline, hence, the sightline-averaged fractional charges will vary as well. The two GC sightlines, which are quite close to one another are much more similar to each other, as might be expected.

\subsection{PAH Size and  Aromatic Carbon Abundance}

To guide our analysis of PAH size and aromatic carbon abundance, we have selected the compact PAH circumovalene, C$_{66}$H$_{20}$, as the parent carrier for a small family of similar-size, compact PAHs \citep[see compact PAH images in Figure 3 in][]{ric12}.   This choice was motivated by four factors: (1) a theoretical prediction of the wavelength of its peak UV extinction feature is consistent with observations \citep{ste11}; (2) based on laboratory measurements of the temperature-dependent PAH IR absorption spectrum, the predicted interstellar $3.3~\mu$m emission feature from circumovalene is consistent with the observed interstellar peak wavelength and width \citep{tok21}; (3) based on many IR emission band ratio studies, the average PAH size generally falls within the interval N$_{\rm{C}}$ = $50-100$ \citep[][and references therein]{kni21};  and (4) in model fits to several Class A emission spectra \citep{die04}, C$_{66}$H$_{20}$, in various charge and isomeric configurations (e.g., the 5-7 ring defect and two side-by-side six-member aromatic rings), made the largest contribution among the compact PAHs included in each fit (see Table 3 in \citet{ric12} and Table 3 in \citet{and15}).

 The precise boundaries of the size interval are still under debate, as is its associated size distribution and its  composition (i.e., irregular vs. compact PAHs).  Based on the previous discussion, we posit that the PAH size interval is approximately  50 $<$ N$_{\rm{C}}$ $<$ 100, with the most abundant parent C$_{66}$H$_{20}$, and a FWHM of order $\Delta$N$_{\rm{C}}$ $\sim$ 25.  We show later that the compact PAHs are the dominant, but not sole, carriers of the UV bump and the $3.2-3.3~\mu$m absorption, with a sizeable contribution from irregular PAHs.  From Figure 3 in \citet{ric12} there are 18 neutral, compact PAHs in the above size interval, but only 8 within the FWHM.  On the other hand, there is a much larger number, $\sim$100, of neutral, irregular PAHs (see the NASA PAH Database).  Each compact and irregular neutral PAH has three charge states and can also host a few extra hydrogen atoms (i. e., be superhydrogenated).  Given the enormous number of potential PAH carriers, it is a challenge to understand how a small number of compact PAHs can account for both the IR emission and absorption bands and the UV absorption spectra.

The idea that a small number of compact PAH carriers could account for the IR emission bands originated from \citet{ric12}, was further quantified by \citet{and15}, who dubbed it the grandPAH hypothesis, and has received recent attention \citep{tok21,mac22,cho23}. \citet{and15} demonstrated that $\sim$12 different PAHs can account for roughly 80\% of the PAHs used to fit the  Class A IR emission spectrum, with irregular PAHs accounting for around 17\% of the PAHs. As mentioned earlier, C$_{66}$H$_{20}$ and its charge and structural isomers are the largest contributor of carbon to the model fit, roughly 27\%. 

  Our simplified analysis of the size distribution offers a partial rationalization of the grandPAH hypothesis, and is in  general agreement with the more quantitative modeling of \citet{and15} We have found that as few as $\sim$8 neutral, compact PAHs, with their associated charge states and with C$_{66}$H$_{20}$ as the most abundant parent, may be the primary carriers of the IR absorption and emission bands.  While there appears to be a significant contribution by irregular PAHs, it is likely to be spread over hundreds of different irregulars; thus, it will manifest as a broad pseudo-continuum contributing to the emission plateau, and not as narrow spectral features.  On the other hand, the large, compact PAHs tend to have similar, overlapping features and, as is seen in IR emission, they combine to produce distinct spectral features.  We  determine below that the carbon abundances in compact PAHs are roughly a factor of three times greater than that in the irregular PAHs, and note that \citet{ric12} speculated that the greater abundance of compact PAHs may be a consequence of their higher chemical stability.

Below we consider the general viability of a mean, Milky Way aromatic PAH carbon abundance of $\sim$45 ppm, tied to an effective PAH size of N$_{\rm{C}}$ = 66.   This choice is in the middle of the range of previous abundance estimates, which span $30-60$ ppm \citep{dra21a,dra21b,ehr10, tie08, dra07} and, as shown below, it is consistent, to better than a factor of two, with the observed opacities of the $3.2-3.3~\mu$m absorption and the UV absorption feature.  
 
\subsubsection{Estimated Carbon Abundance from the UV Absorption Feature}

For the average Milky Way sightline, the peak optical depth of the 2175 \AA\ absorption feature, observed through a visual extinction, $A_{V}$, of 1 mag (corresponding to a H atom column density, $N$(H), of 2.2 $\times$ 10$^{21}$ cm$^{-2}$, is  $\tau_{\rm p}$ $\approx$~1.25  \citep{car89,ste11}, after subtraction of the underlying dust continuum optical depth at that wavelength. The DFT methodology used by \citet{ste11} appears to produce accurate wavelengths for the UV bump, as is evident in their comparison (their Figure 1) to the laboratory data for a thin film of the PAH hexa-peri-benzocoronene  (HBC, C$_{42}$H$_{18}$). We assume that the predicted wavelength for C$_{66}$H$_{20}$  is comparably accurate. Assuming C$_{66}$H$_{20}$  as the characteristic carrier, theoretical calculations by \citet{ste11} produce a peak UV bump absorption cross section, C$_{\rm abs,p}$, of approximately  1.2 $\times$ 10$^{-17}$ cm$^{2}$ per carbon atom. Laboratory UV absorption measurements of ovalene, C$_{32}$H$_{12}$, yield a comparable peak cross section, allowing for some broadening to match the width of the UV bump \citep{lea95}. The values of the above parameters result in  a carbon abundance, [C/H] = $\tau_{\rm p}$/$N$(H)C$_{\rm abs,p}$ = 4.7 $\times$ 10$^{-5}$ (47 ppm), with an estimated uncertainty of  $\sim$ $\pm$25\%. This is in good agreement with our previously assumed value of 45 ppm. We expect that this calculation leads to a fairly accurate abundance estimate, because the peak absorption cross section per carbon atom is nearly independent of size and charge state \citep{mal04}. For comparison, in a recent theory-based modeling study of the UV bump, \citet{lin23} derive a similar aromatic carbon abundance, 40 ppm.

 \begin{table}
\begin{center}
\caption{Carbon Abundance Estimate toward 2M1747 for Compact C$_{66}$H$_{20}$- and Irregular C$_{66}$H$_{24}$-related PAHs$^a$}
\begin{tabular}{ccccc}
\hline\hline
Species & S$^{b,c,d}$ & $\Delta\nu^{e}$ & $\tau_{\rm p}^{f,g}$ & [C/H] \\
& km mole$^{-1}$ & FWHM cm$^{-1}$  & $\times$ 100 & ppm \\
\hline
C$_{66}$H$_{20}$$^{+}$ & 351 & 23.0 & $3.46 - 0.25$ & 22.0 \\
C$_{66}$H$_{20}$$^{0}$ & 686 & 29.6 & $2.69-0.38$ & 10.7 \\
C$_{66}$H$_{20}$$^{-}$ &1097 & 27.9 & $3.98-0.96$ & 8.2 \\
C$_{66}$H$_{24}$$^{+}$ & 99 ~~274 &  23.0 ~~21.4  & 1.87/3 & 6.3 \\ 
C$_{66}$H$_{24}$$^{0}$ & 190~~200 & 29.6 ~~21.4 & 1.87/3 & 6.3 \\
C$_{66}$H$_{24}$$^{-}$ & 452~~200 & 27.9 ~~21.4 & 1.87/3 & 6.3\\
\hline\hline
\end{tabular}
\end{center}
$^{a}$ For diffuse cloud visual extinction of 30 mag \citep[see][and text]{geb21} \\
$^{b}$  Multiply S by 10$^{5}$/6.022$\times$10$^{23}$ for cm molecule$^{-1}$ \\
$^{c}$ Values from NASA PAH Database \\
$^{d}$ Left-hand S values in column 2 for irregular PAH bands for $\nu$ $<$ 3100 cm$^{-1}$ ($\lambda$ $>$  3.226 $\mu$m)  that overlap with compact PAH bands; right-hand S values for irregular PAHs bands for $\nu$$>$ 3100 cm$^{-1}$ to the blue of the compact PAH interval.
 \\
$^{e}$ Values from Table 1 for the compact PAHs.  The irregular bands (left-hand values) that overlap the compact PAHs are assumed to have the same widths as for the compact PAHs.  The irregular PAHs (right-hand values) to the blue of the compact PAHs are assigned the width found for the irregular blue feature in Table 1. \\
$^{f}$ Values from Table 1, except irregular contributions to the compact PAH values are subtracted from the Table 1 values\\
$^{g}$ Assumes three charge states of C$_{66}$H$_{24}$ have equal abundances
\end{table}

\subsubsection{Estimated Carbon Abundance from the $3.2-3.3~\mu$m Absorption}

Here we quantify the total aromatic carbon abundance tied up in PAH clusters.  We first place PAH clusters within the broader context of interstellar dust, since that guided our approach to calculating the aromatic carbon abundance. The chemical composition, charge distribution, multi-component structures, and sizes of the interstellar dust grains are still active areas of research (see the recent review by \citet[][and references therein]{her22}.  To guide our approach, we considered the dust model of \citet{hen23} that represents interstellar dust in terms of two separate components, PAH-like nanoparticles and astrodust (which covers the larger, amorphous, carbonaceous and siliceous grains).  Our PAH clusters, described in the next section, are analogous to the \citet{hen23} PAH-like nanoparticles.  They are modeled in terms of two size components, peaking at radii of $\sim$6.5~\AA\ and $\sim$45~\AA\ \citep[see Fig. 1 in][]{hen23}.  The smaller size distribution pertains to the IR emission, and the larger to UV absorption \citep{dra21a}. The nanoparticle model has yet to be applied to the interstellar IR absorption observations, but one would expect the emission and absorption carriers to be closely related.  While we are not explicitly applying the \citet{hen23} model to the IR absorption, the model provides a general framework to rationalize our approach, presented below, to estimate the carbon abundances from PAH clusters in absorption.
 
 For simplicity and because they represent the majority of the PAH carbon abundance, we focus on the smaller PAH nanoparticles whose peak size of 6.5~\AA\ corresponds to N$_{\rm{C}}$ $\sim$132 \citep[see Eq. 21 in][]{hen23}. This effective PAH size corresponds to two of our parent PAH, C$_{66}$H$_{20}$ (i.e., a dimer).  The emission occurs at an earlier stage of the evolution of the PAH size distribution and favors gas phase PAHs, whereas the absorption involves PAH molecules that have been injected into the interstellar medium (ISM) and favor larger PAH clusters formed by continued coagulation. The coagulated PAH clusters may also develop mantles of non-aromatic hydrocarbons \citep{chi13}.  As discussed later, an average cluster injected into the ISM consists of $\sim$5 PAH monomers (i.e., N$_{\rm{C}}$ $\sim$330).  \citet{hen23} estimate that approximately 47\% of the PAHs for N$_{\rm{C}}$ $\sim$ 330 will be ionized in the diffuse ISM.  Thus, we expect a significant fraction of charged PAH clusters.  Since each of the PAH molecules in a cluster can carry a charge, a given cluster may carry a few charges.  We note that \citet{dra21b} estimate a total aromatic carbon abundance of $\sim$40 ppm.
  
Below we estimate the aromatic carbon abundances of the PAHs that we have identified with the Gasussian components fitted to the observed spectra in Figures 4-6. We assume that the compact PAH C$_{66}$H$_{20}$ in its three charge states is representative of the three absorption features in the 3.23-3.31$~\mu$m interval, and that the irregular PAH C$_{66}$H$_{24}$ in its three charge states are representative of the absorption feature observed near 3.21 $\mu$m.  Irregular PAHs in their three charge states also have absorption features that coincide with the three compact PAH features.  As mentioned earlier, both the compact and irregular PAHs display solo and duo CH(S) bands that occur in the above compact PAH interval, whereas the irregular PAHs also have trio and quartet CH(S) bands that occur to the blue of the compact PAH interval.

The expression for computing the carbon fractional abundance, [C/H], for a specific Gaussian component is identical to that above for the UV bump. The peak optical depths are based on the Gaussian fits (see Table 1), and the peak absorption cross sections are estimated from theoretical band strengths, S, and the  band widths,  $\Delta \nu$ (FWHM), from the Gaussian  fits (see Table 2).  The peak absorption cross section per carbon atom is computed from C$_{\rm abs,p}$ = $f_{\rm G}$$f_{\rm cor}$S/(N$_{\rm C}$$\Delta \nu$), where $f_{\rm cor}$ approximately  corrects the theoretical DFT band strength to experimental data, and $f_{\rm G}$ = 2(ln2/$\pi$)$^{0.5}$ is the normalization factor for the Gaussian band shape used in the spectral fits.  The total carbon abundance is the sum of the [C/H] over all of the Gaussian fit components. We apply an estimated correction factor of $f_{\rm cor}$ = 0.6 to all the theoretical aromatic CH(S) S values derived from the NASA PAH Database. For example, the experimental band strengths for pyrene (C$_{16}$H$_{10}$) and coronene (C$_{24}$H$_{12}$) are factors of 0.5 and 0.7, respectively, less than their DFT values. For later use, we note that the expression for [C/H] can be rearranged to yield the peak optical depth, $\tau_{\rm p}$ = [C/H]$N$(H)C$_{\rm abs,p}$ when the carbon abundance is known.  We use this to determine the optical depths of the irregular bands  that reside in the compact PAH interval.  For them the abundance is that determined for the irregular feature that falls outside of the compact PAH interval.  The derived irregular optical depths are subtracted from the total optical depths from the Gaussian fits, in order to estimate the carbon abundances specifically for the compact PAHs.
 
The expression for [C/H] is equivalent to that derived from the UV absorption (see Section 3.3.1), taking into account the relationship C$_{\rm{abs,p}}$ = $f_{\rm{G}}$S / (N$_{\rm{C}}$$\Delta\nu$) for the UV bump. The expression for [C/H] can also be rearranged to yield the peak optical depth  $\tau_{\rm p}$ =  [C/H]$f_{\rm{G}}$$f_{cor}$S$A_{V}$$N$(H)/(N$_{\rm{C}}\Delta\nu$), to determine the optical depths of the irregular bands that reside in the compact PAH interval. Those optical depths need to be subtracted from the total (i.e. from the Gaussian fit), in order to estimate the carbon abundance in compact PAHs. The above expression pertains to sightlines of arbitrary visual extinction, $A_{V}$, where $N$(H) = 2.2 $\times$ 10$^{21}$ cm$^{-2}$ when $A_{V}$ = 1 mag.
 
The values of the parameters used in the estimate of the carbon abundance for 2M1747 are displayed in Table 2.  The diffuse cloud values of $A_{\rm{v}}$ are $\sim$30 mag \citep[][see the Introduction]{geb21}, 29 mag \citep{chi13}, and 10 mag \citep{hen20} for 2M1747, QC, and Cyg OB2-12, respectively.  The values of S for the irregular PAHs are separated into the two components corresponding to the bands within, and to the blue, of the compact PAH interval.  The bands to the blue have similar S values, where, for simplicity, we used the average of the three bands, S = 226 km mole$^{-1}$.  Given the similarity of the three S values, making different assumptions in regard to the relative contribution of each charge component will not produce meaningful differences in the distribution-averaged S value.  We note that the S values for the irregular PAHs in the compact PAH interval are roughly a factor of three lower than their compact PAH counterparts.  

We first determine the carbon abundance for the irregular blue component, [C/H] = 6.3 ppm for each of the assumed three overlapping bands, for a total aromatic carbon abundance of $\sim$19 ppm.  Given these abundances, we can estimate and remove the contribution of the irregular PAHs to each of the compact PAH peak optical depths.  We can then compute the carbon abundances for the ``corrected" compact PAH features.  The final 2M1747 results are [C/H] = 19 and 41 for the irregular and compact PAHs, respectively, for a total aromatic carbon abundance of 60 ppm.  For QC, [C/H]  = 11 and 48 for the irregular and compact PAHs, respectively, for a total aromatic carbon abundance of 59 ppm. For Cyg OB-12, [C/H] = 11 and 36 for the irregular and compact PAHs, respectively, for a total aromatic carbon abundance of 47 ppm. 

The uncertainties in the values of several of the parameters that enter into the IR abundance calculations are not well established.  We estimate that the uncertainty in the carbon abundances is {\it at least} 25\% and that therefore our estimates based on IR data are consistent with the UV-derived abundance of 47 ppm.  The UV value is an average for the entire Milky Way, whereas the IR values are for three specific sightlines, which may deviate from the Galaxy average.  An additional source of uncertainty is the choices of representative PAHs.  However, the band strengths scale roughly linearly with N$_{\rm{C}}$, which implies that N$_{\rm{C}}$/S is approximately independent of the assumed molecular size. Thus, quantities, such as [C/H] and $\tau_{\rm p}$, that depend on N$_{\rm{C}}$/S are insensitive to the sizes of the representative PAHs.

An interesting result of our IR analysis is the significant contribution of irregular PAHs.  We find that the irregular contribution accounts for 19-32\% of the total aromatic carbon abundance.  The irregular feature coincides with the weakest portion of observed absorption, but in our deconvolution it appears at comparable positions, widths, and relative intensities in all three sources (see Table 1); hence, it is likely real.  While compact PAHs may be the dominant carrier type, it makes chemical sense to also have a significant irregular PAH carrier component.  It is difficult to imagine a PAH chemical formation route that only produces compact PAHs.

\begin{table}
\begin{center}
\caption{6.2~$\mu$m Optical Depth Estimates toward Quintuplet Cluster for Compact C$_{66}$H$_{20}$-related and Irregular C$_{66}$H$_{24}$-related PAHs}
\begin{tabular}{cccccc}
\hline\hline
Species & S$^{a,b}$ & $f_{cor}~^{c}$ & [C/H]$^{d}$ & $\Delta\nu^{e}$ & $\tau_{\rm p}$ $\times$ 100 \\
& km mole$^{-1}$ & & ppm & FWHM cm$^{-1}$ & \\ 
\hline
C$_{66}$H$_{20}$$^{+}$ & 630 & 0.35 & 20 & 40 & 1.64 \\
C$_{66}$H$_{20}$$^{0}$ & 33 & 0.49 &14 & 40 & 0.077 \\
C$_{66}$H$_{20}$$^{-}$ & 362 & 0.35 & 14 & 40 & 0.68 \\
C$_{66}$H$_{24}$$^{+}$ & 1194 & 0.35 & 3.8 & 40 & 0.60 \\ 
C$_{66}$H$_{24}$$^{0}$ & 54 & 0.49 & 3.8 & 40 & 0.037 \\
C$_{66}$H$_{24}$$^{-}$ & 804 & 0.35 & 3.8 & 40 & 0.40 \\
\hline\hline
\end{tabular}
\end{center}
$^{a}$ Multiply S by 10$^{5}$/6.022$\times$10$^{23}$ for cm molecule$^{-1}$ \\
$^{b}$ Values from NASA PAH Database for  $1551-1590$ cm$^{-1}$ ($6.289-6.447 \mu$m)\\
$^{c}$ Estimated from data-theory comparison for a small PAH (pyrene)\\
$^{d}$ Estimated from analysis of $3.2-3.3~\mu$m absorption (this paper) \\
$^{e}$ Observed FWHM \citep{chi13} \\
\end{table}

\subsubsection{Comparison with the 6.2 $\mu$m interstellar band}

Although the focus of this paper is on the $3.2-3.3~\mu$m CH(S) absorption, we note that the absorption spectra for a number of bands longward of 5~$\mu$m have been measured toward both QC and Cyg OB2-12 \citep{chi13,hen21}. We consider the Gaussian fit component at 6.25 $\mu$m (1600 cm$^{-1}$) in \citet[][red curve in their Figure 2]{chi13} that is associated with the aromatic CC stretch. The observed width and peak optical depth of this feature are 40 cm$^{-1}$ and 0.30, respectively. From a comparison of the measured and theoretical band strengths for the pyrene (same D2h symmetry as C$_{66}$H$_{20}$) cation and neutral, we estimate  correction factors $f_{cor}$ of  $\sim$0.35 and $\sim$0.49, respectively.  Experimental data are not available for the anion, for which we assume the same correction factor as for the cation. We assume the correction factors for the irregular PAH, C$_{66}$H$_{24}$, are the same as its C$_{66}$H$_{20}$ charge counterparts. We consider all the bands of the cation, neutral, and anion, for the irregular and compact PAHs, that fall within the 40 cm$^{-1}$ FWHM interval about the peak. The DFT calculations for the pyrene cation place its band center 29 cm$^{-1}$ lower ($\sim$0.11 $\mu$m longer wavelength) than the laboratory measurements. Applying this correction implies that the contributing bands in the NASA PAH Database fall within the 1551-1590 cm$^{-1}$ ($6.289-6.447 \mu$m), interval; i.e. in the 40 cm$^{-1}$ Gaussian width interval of the DFT calculations.  We calculate the peak optical depths for all of the contributing bands of the irregular and compact representative PAHs in order to compare their sum to the observed peak optical depth.  The optical depth calculations are based on the previously-derived carbon abundances for the $3.2-3.3~\mu$m absorption.  The peak optical depths are calculated using the previously presented expression for $\tau_{\rm p}$. The results are displayed in Table 3.

The total model-based peak opacity, 0.034, is in acceptable accord with the measured value of 0.030. Thus, it appears that for QC, the $3.2-3.3~\mu$m and 6.2~$\mu$m absorption features are consistent with the same total aromatic carbon abundance of 59 ppm, , as well as the individual carbon abundances for each of the six component bands. 

\begin{figure}
\begin{center}
\includegraphics[width=0.45\textwidth, angle=-0]{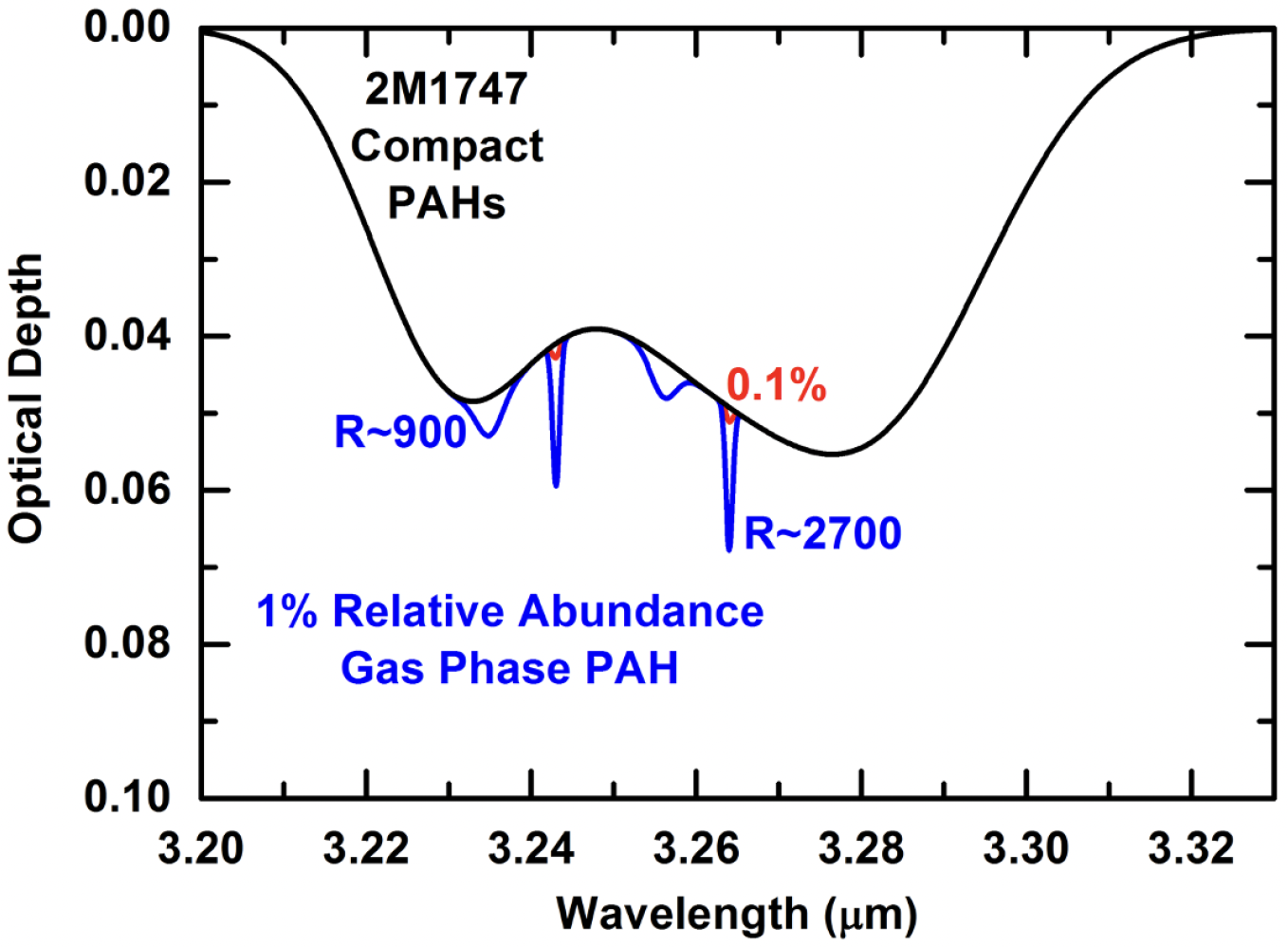}
\end{center}
\caption{Simulations of the absorption by a single gas phase PAH with an abundance of 1\% (blue curve) relative to the total PAH abundance (including PAH clusters) as observed at the current R = 900 and at the highest JWST R = 2700, superimposed on the Gaussian fit to the 2M1747 spectrum in Figure 4 (black curve), The very small red absorption profiles are for a 0.1\% relative abundance.}
\end{figure}

\subsection{Cluster Formation and Spectral Smoothness}

The smoothness of the corrected 2M1747 profile (Figure 3b) suggests that the PAHs are in a condensed phase, such as a grain or molecular cluster. However, as pointed out in the previous section, a gas-phase PAH produces narrow spectral features; these would be detectable if their peak optical depths exceed the noise level.  JWST/NIRSpec higher resolution spectra of the fundamental $3.2-3.3~\mu$m absorption on DIB sightlines might detect such narrow, features, instead or in addition to the broad, smooth features in the spectra analyzed here.  Figure 8 shows how a gas-phase PAH profile would appear at different spectral resolutions, and abundances, perched on top of the Gaussian fit to the 2M1747 spectrum. The gas-phase lines are assumed to be doublets due to resonance-based splitting of the two strongest, degenerate lines \citep{mac18}, as seen in the laboratory spectrum of ovalene, C$_{32}$H$_{14}$ \citep{job94}.

At R = 900 the optical depth of a gas-phase absorption feature from a PAH with an abundance of 1\% relative to the total PAH abundance (including PAHs in clusters) is ~0.004 (Figure 8), which would be borderline discernible from the noise in Figure 6, but readily detectable by James Webb Space Telescope (JWST). The much higher S/N and spectral resolution of the JWST may enable detection of gas phase PAHs at the $\sim$0.1\% abundance level (see small red features in Figure 8).  However, the following cluster formation analysis suggests that the total abundance of gas phase PAHs is $<$1\%, implying that the abundance of any specific gas-phase PAH is $<$0.1\%.

We assume that PAHs are formed in the outflows of post-AGB stars, whereupon they form PAH molecular clusters and are subsequently injected into the diffuse ISM \citep{leq90,chi98,che11}. \citet{chi98} proposed that the presence of the 3.4~$\mu$m absorption feature in a post-AGB star is the result of the evolution of gas-phase monomer PAHs into PAH clusters, bound together by aliphatic chains during their residence within the outer envelopes of the post-AGB star. We present here a quantitative model for this evolutionary process.  Our focus is on the neutral molecular envelope, which is bound by the inner \ion{H}{2} and outer stellar wind regions.  We do not expect significant cluster formation to occur in \ion{H}{2} regions because (1) their radiation fields include far UV photons capable of dissociating PAH dimers, and (2) the gas kinetic temperatures of $\sim$10,000 K should result in small sticking coefficients for PAH-PAH collisions \citep{tot12} ).  If the assumption of no  dimer formation in the \ion{H}{2} envelope is incorrect, then we will have underestimated the already substantial monomer depletion estimated. Additionally, we have not considered collisions between free-flying PAHs and aliphatic particles in the AGB envelopes and ISM, which will further reduce the small fraction of free-flying PAHs.

We employ a simplified model (see Appendix B for the model derivation) for monomer depletion and cluster formation that is sufficient to establish their order of magnitude abundances.   We divide the neutral envelope into two sub-regions.  The first is dominated by monomer depletion due to dimer formation.  The second is dominated by monomer depletion due to trimer, and large cluster, formation.  We estimate that the mean cluster size injected into the ISM is composed of around five PAHs and tails off at the small and high sizes, dimers and decemers, respectively.  The physical properties used for estimating the depletion and formation rates are based on the post-AGB parameter values for NGC 7027 \citep{has00}.

The total monomer PAH depletion factor for the neutral envelope is given by the product of the depletion factors (see Appendix B) for the two regions, 0.11 $\times$ 0.41 = 0.045.  An approximate consideration of the effects of the outer stellar wind shell suggests that further monomer depletion of around 0.8 occurs, resulting in a final monomer depletion factor of $\sim$0.036.  This depletion relates to PAHs that are just entering the diffuse ISM. Further monomer depletion and cluster growth may occur during the long residence time in the ISM.  We have not assessed these effects; consequently, our estimates are lower bounds.  An addition consideration is the different destruction time scales for the monomers vs. clusters.  A C$_{66}$H$_{20}$ monomer will survive for $\sim$40 Myr as compared to its clusters, which will survive for $\sim$120 Myr \citep{mic11}.  This means that in the steady state of formation and destruction, the cluster PAH abundances will increase over those for the monomers by another factor of $\sim$120/40 = 3.  This lowers the monomer depletion factor to 0.036/3 = 0.012.

 The depletion factor of 0.012 applies to the total PAH monomer population.  However, there are a variety of PAH sizes, charge states, superhydrogenation levels, etc. defining specific PAHs.  Assuming that there are a few, say three, PAH sizes, three charge states, and two superhydrogenation levels (i.e., 0 or 1 extra H), then there are eighteen unique monomers, each with a depletion factor around 0.012/18 = 0.0007.  This implies that the narrow features of specific monomer PAHs will be very difficult to observe superimposed on the broader, cluster-dominated absorption spectrum.
  
\section{ SUMMARY}

  The key results from this study are as follows:

\begin{enumerate}

\item The initially reduced and published 2M1747 ground-based spectrum of the $3.2-3.3~\mu$m interstellar absorption exhibits strong, residual, telluric spectral features, due to the airmass difference between the source and the standard star sightlines.  The initially reduced spectrum has been corrected by the addition of its associated airmass residual optical depth spectrum.

\item The corrected absorption spectrum of 2M1747, and previously published spectra of the Quintuplet Cluster and of Cyg OB2-12 in the $3.2-3.3~\mu$m interval, neither of which require correction, were each fit well by superpositions of four Gaussians.  The central wavelengths and widths of the Gaussians in each source are closely similar, but their peak optical depths vary more.  The three longest-wavelength Gaussians fall within the range $3.23-3.31~\mu$m, which coincides with the wavelengths of the mono and duo fundamental CH stretch (CH(S)) absorptions by positive, neutral, and negatively charged, compact and irregular PAHs.  The shortest wavelength Gaussian, near $3.21~\mu$m, is attributed to trio and quartet CH(S) bands in irregular PAHs.

\item The compact and irregular PAHs responsible for the $3.2-3.3~\mu$m absorption are mainly in condensed phases, most likely in modest-sized clusters composed of ten or fewer PAHs.  We have presented a  cluster formation model that supports the idea that the PAHs arise in the outer envelopes of post-AGB stars\, prior to ejection into the diffuse interstellar medium.  We estimate that $>$99\% of the aromatic carbon is tied up in clusters, with 0.1\%, or less of the carbon in the form of any specific, gas-phase, monomer PAH.

\item A compact, C$_{66}$H$_{20}$-based PAH family with a mean Milky Way carbon abundance of $\sim$45 ppm is consistent with the wavelength and optical depth of the 2175~\AA~absorption feature, the width and wavelength shift of the most abundant Class A $3.3~\mu$m emission feature and the average optical depth of the $3.22-3.31~\mu$m and the 6.2~$\mu$m interstellar absorption features. The irregular feature at $\sim$3.21~$\mu$m contributes an additional, source-averaged, aromatic carbon abundance of $\sim$14 ppm. 

\end{enumerate}

  This study supports a version of the PAH hypothesis wherein (1) compact PAHs in the 
50 $\lesssim$ N$_{\rm{C}}$ $\lesssim$ 100 size range, with a most abundant parent PAH in the vicinity of C$_{66}$H$_{20}$, are the primary, but not sole, carries of the IR absorption and emission bands as well as the UV 2175~\AA~absorption bump; (2) irregular PAHs account for approximately 25\% of the total aromatic carbon abundance; and (3) the IR and UV absorption features arise predominantly from PAH molecular clusters. This study is qualitatively consistent with the grandPAH hypothesis, in that there may be a relatively small number of  compact PAH carriers: roughly 10 or fewer neutrals, with their associated charge states.  Additionally, we cannot rule out contributions from ring defects and superhydrogenated isomers. JWST spectroscopy of  the $3.2-3.3$ $\mu$m region on additional diffuse cloud sightlines should provide stringent tests of our analysis and our interpretation of the $3.2-3.3$ $\mu$m absorption band. 

\acknowledgments

We thank Alan Tokunaga for many helpful discussions during the course of this project and for comments on the draft of this manuscript.  We thank Alexander Berk for providing the MODTRAN atmospheric transmittance spectra.  We also are grateful to Aigen Li, Lou Allamandola, Greg Sloan, and Kathleen Kraemer for helpful technical discussions and to the referee for helpful comments and suggestions. This research is based in part on observations obtained at the international Gemini Observatory (reference GN-2016B-FT-12), a program of NSF's NOIRLab, which is managed by the Association of Universities for Research in Astronomy (AURA) under a cooperative agreement with the National Science Foundation on behalf of the Gemini Observatory partnership: the National Science Foundation (United States), National Research Council (Canada), Agencia Nacional de Investigaci\'{o}n y Desarrollo (Chile), Ministerio de Ciencia, Tecnolog\'{i}a e Innovaci\'{o}n (Argentina), Minist\'{e}rio da Ci\^{e}ncia, Tecnologia, Inova\c{c}\~{o}es e Comunica\c{c}\~{o}es (Brazil), and Korea Astronomy and Space Science Institute (Republic of Korea).


\facilities{Gemini:Gillett, United Kingdom Infrared Telewscope, ISO}

\clearpage
\appendix

\section{Model for the Airmass Residual Optical Depth}

 We derive an expression for the airmass residual optical depth, $\Delta\tau_{air}$, that corrects the initially-reduced absorption spectrum for a difference in airmass between the source and standard star.  The transmittance on a sightline through the Earth's atmosphere, $t_{air}$, can be expressed in terms of two optical depth components,
\begin{equation}
{\rm{-ln}}(t_{air}) = \tau_{air} = \tau_{line} + \tau_{cont}, 
\end{equation}				     

\noindent where $\tau_{air}$ is the total optical depth, $\tau_{line}$ is the interval-averaged optical depth due to molecular lines (H$_{2}$O and CH$_{4}$ for the $3.2-3.3~\mu$m interval), and $\tau_{cont}$ is the smooth and slowly varying optical depth from continuum absorption \citep[see][]{mia12}.  The continuum arises from two main contributions, the far wings of strong H$_{2}$O lines and water dimers. As discussed below, the reason for the separation into two components is that they scale differently with respect to a {\it fractional change} in path length (i.e., airmass), denoted by $\delta_{\rm{air}}$ (i.e., the difference in airmass divided by the airmass). For 2M1747 and its standard star, the airmass difference was 0.06, and the standard star airmass was 1.58, yielding    $\delta_{\rm{air}}$ = 0.06/1.58 = 0.038. The two components can be determined from the measured standard star telluric transmission spectrum.  For 2M1747 $\tau_{cont}$ $\approx$ 0.05 (see Figure 2), which is discernible the narrow, highly-transmissive regions between highly-absorbing intervals.  Given $\tau_{cont}$, one can determine $\tau_{line}$ from equation A1 (i.e., $\tau_{line} = \tau_{air} - \tau_{cont}$).  For a change in airmass, the modified, total optical depth can be expressed as,

\begin{equation}
\tau'_{air} = (1 + \delta_{air})^{f_{RT}} \tau_{line} + (1 + \delta_{air})\tau_{cont},
\end{equation}

\noindent where $f_{RT}$ is a radiation-transfer (RT) factor that depends on the type of line shape (i.e., Lorentzian, Doppler, or Voigt) and the effective peak opacity for the lines in each spectral interval, as defined by the spectral resolution \citep[for details see the discussion of band models in][]{lud73}, For the current problem, the line shape corresponds to a pressure-broadened Lorentzian, where $f_{RT}$ may vary for each spectral interval, and falls within the range of  0.5 $\le$ $f_{RT}$ $\le$ 1.0. The lower limit, 0.5, is referred to as the strong line limit and corresponds hereto opaque (or nearly so) lines (i.e., $t$ $<$ 0.1).  The upper limit, 1.0, is referred to as the weak line limit (equivalent to Beer's Law) and corresponds here to a constant, continuous, and small optical depth across the spectral resolution interval.  For small airmass differences, $\delta_{\rm{air}}\ll$1, as is the case here, the corresponding airmass optical depth residual can be derived from equation A2 as,

\begin{equation}
\Delta\tau_{air} = f_{RT} \delta_{air} \tau_{line} + \delta_{air} \tau_{cont}
\end{equation}

\noindent where for reasons discussed earlier, $f_{RT}$ is treated as a wavelength-independent fitting parameter (see discussion of Figure 4 in Section 2.1).  For the $3.2-3.3~\mu$m spectrum of 2M1747 we determined that $f_{RT}$ = 0.79$\pm\ 0.13$, which results in  $\delta_{air}f_{RT}$ = 0.030. The airmass difference, $\delta_{air}$, is determined by the measurement geometry.  Its inclusion in equation A3 assumes that both the H$_{2}$O and CH$_{4}$ column amounts varied only with airmass.  Since CH$_{4}$ is uniformly mixed, zenith angle scaling will apply.  The H$_{2}$O mixing ratio can vary both spatially and temporally.  If the airmass and H$_{2}$O mixing ratio have different fractional variations, one would set $\delta_{air}$ equal to the H$_{2}$O fractional difference, because H$_{2}$O is almost always the dominant absorber in the $3.2-3.3~\mu$m interval, even on a dry site such as Maunakea. However, in this instance the point is moot, because the H$_{2}$O column did not vary during the observations.

\section{Model for PAH Monomer Depletion and Cluster Formation}

Here we derive expressions for the depletion of gas-phase monomer PAHs and the corresponding formation of PAH clusters in the outflow of post-AGB stars.  The PAH clusters discussed here are composed of van der Waals-bonded monomer PAHs, which, for a C$_{66}$H$_{20}$ parent PAH pair, are sufficiently strongly bound , $\sim$3.4 eV, to avoid UV photo-dissociation in the stellar envelopes and ISM enviroments\citep{lan21}. We focus on the neutral envelope of the post-AGB flow and briefly consider the outermost flow envelope, which deposits gas and dust into the diffuse ISM.  We simplify the model by dividing the neutral envelope into two sub-regions.  The first is dominated by monomer depletion due to dimer formation.  The second is dominated by monomer depletion due to trimer and quatromer formation.  The physical properties used for estimating the depletion and formation rates are based on the post-AGB parameter values for NGC 7027 \citep{has00}.
  
  The governing equation for monomer depletion in region 1 is,
  
 \begin{equation}
 \frac{dn_{m}}{dt} =  -2n_{m}^{2}\sigma_{m,m}v_{m,m}
  \end{equation}
 							                                
\noindent where $n_{m}$ is the monomer number density, $\sigma_{m,m}$ is the monomer-monomer collision cross section, and $v_{m,m}$ is the average collision velocity.  The factor of 2 arises because each dimer that is formed removes two monomer PAHs from the gas phase.  We assume that the sticking coefficient for dimer formation is unity \citep{mon14}. The initial monomer density, $n_{m}$(t=0), is given by,

\begin{equation}
n(0)_{m} = \frac{\rm [C/H]}{\rm{N}_{\rm{C}}}n_{\rm{H}}
\end{equation}

\noindent where [C/H] (= 4.5 $\times$ 10$^{-5}$ as discussed in Section 3.3) is the fractional abundance of aromatic carbon atoms, N$_{\rm{C}}$ is the number of carbon atoms in the characteristic PAH, circumovalene (C$_{66}$H$_{20}$), and $n_{\rm{H}}$, 8.1 $\times$10$^{6}$ cm$^{-3}$ for NGC 7027 \citep{has00}, is the hydrogen number density in region 1. These parameter values give an initial monomer PAH density of $n$(0)$_{m}$ = 5.5 cm$^{-3}$. The collision cross section is obtained from

\begin{equation}
\sigma_{m,m} = \pi(2r_{m})^{2}
\end{equation}
				                  
\noindent where $r_{m}$ is the effective radius of the slightly-elliptical, characteristic PAH.  For C$_{66}$H$_{20}$, $r_{m}$ = 6.0~\AA, which results in a collision cross section $\sigma_{m,m}$  = 4.5 $\times$ 10$^{-14}$ cm$^{2}$.  The average collision velocity is calculated from

\begin{equation}
v_{m,m} = (\frac{8k_{\rm{B}}T\rm{N}_{\rm{avo}}}{\pi\mu})^{0.5}, 
\end{equation}

\noindent where $k_{\rm{B}}$ is Boltzmann's constant, T is the temperature (800 K for NGC 7027), N$_{\rm{avo}}$ is Avogadro's number, and $\mu$ (0.5m$_{\rm{PAH}}$ = 406 amu for C$_{66}$H$_{20}$) is the reduced mass for the monomer-monomer collision, yielding $v_{m,m}$ = 2.0 $\times$ 10$^{4}$ cm s$^{-1}$.  There are two controlling time scales to consider.  One is the time scale for traversing region 1,

\begin{equation}
t_{1} = \frac{\Delta R_{1}}{v_{w}}
\end{equation}

\noindent where $\Delta R_{1}$ is the radial thickness of region 1, and $v_{w}$ is the stellar wind velocity (15 km s$^{-1}$ for NGC 7027).  The radial thickness of the neutral envelope is $\Delta R_{1}$ + $\Delta R_{2}$ = 6.8 $\times$ 10$^{-4}$ pc for NGC 7027, where the relative thicknesses are in a ratio of about 0.6/0.4.  The border between the two regions is defined by a monomer depletion of $\sim$90\%.  For NGC 7027 the travel times across the two regions are $t_{1}$ = 8.4 $\times$ 10$^{8}$ s and $t_{2}$ = 5.6 $\times$ 10$^{8}$ s.  

The collision time scale (the inverse of the initial monomer-monomer collision rate) is

\begin{equation} 
t_{m.m} = [2 n(0)_{m}\sigma_{m,m}v_{m,m}]^{-1}
\end{equation}						        

\noindent For region 1 in NGC 7027, this is 1.0 $\times$ 10$^{8}$ s.  The monomer depletion factor for dimer formation in region 1 is given by the integration of equation B4,

\begin{equation}
\frac{n(t_{1})_{m}}{n(0)_{m}} = \frac{1}{1 + \frac{t_{1}}{t_{m,m}}},
\end{equation}

\noindent which, for NGC 7027, is 0.11.  This means that 0.89 of the monomer population is predominantly tied up as dimers, and some larger clusters, by the time $t_{\rm{1}}$ is attained.
 
In order to estimate the monomer depletion time scale in region 2 due to monomer-cluster collisions, we need to estimate the average cluster size in region 2.  The characteristic time scale for cluster size doubling can be approximated by using equation (B9).  This requires a modification for the initial cluster size density, which, for doubling of dimers to form a quatromer, is given by $n$(0)$_{d}$ =  $n$(0)$_{m}$/2).  We estimate that two to three doublings occur in region 2, corresponding to clusters comprised of four (quatromer) to eight (octamer) monomer PAHs.  We take the quatromer cluster as representative of the average size in region 2.  This indicates that the average size of clusters entering the diffuse ISM contain approximately six PAH monomers.

The governing equation for monomer depletion in region 2 due to monomer-quatromer collisions, that give rise to pentamers, is

\begin{equation}
\frac{dn_{m}}{dt} = n(t_{1})_{q}\sigma_{m,q},v_{m,q},
\end{equation}
 (i.e., all quatromers)
\noindent where $v_{m,q}$ = (5/8)$^{1/2}$ due to the difference in reduced masses. The value to adopt for the
monomer-quatromer collision cross section is not obvious.  It is bounded on the low end by $\sigma_{m,m}$, which corresponds to a quatromer with perfectly stacked monomers.  On the high end, it is bounded by 2.25$\sigma_{m,m}$, which corresponds to five (quatromer plus monomer PAHs) side-by-side.  We adopt an intermediate value of 1.6$\sigma_{m,m}$.  The time scale for the monomer-quaromer collisions is given by

\begin{equation}
t_{m,q} = [n(t_{1})_{q}\sigma_{m,q}v_{m,q}]^{-1},
\end{equation}

\noindent where $n$(t$_{1}$)$_{q}$ = 0.25$n$(0)$_{m}$ (i.e., all quatromers). The monomer depletion factor for pentamer formation in region 2 is given by integration of eq.(B11),

\begin{equation}
\frac{n(t_{2})_{m}}{n(t_{1})_{m}} = \rm{exp}(\frac{-t_{2}}{t_{m,q}}),
\end{equation}  
					                                                    
\noindent For NGC 7027 $t_{2}$ = 5.6 $\times$ 10$^{8}$ s and $t_{m,q}$ = 6.2 $\times$ 10$^{8}$ s, yielding a monomer depletion factor of exp(-5.6/6.2) = 0.41.

The total depletion factor for the neutral shell is given by the product of the depletion factors for the two regions, 0.11$\times$ 0.41 = 0.045.  Approximate consideration of the outer stellar wind shell suggest that further monomer depletion of around $e^{-0.2}$ = 0.8  occurs, resulting in a final monomer depletion factor of about 0.036 for PAHs exiting the post-AGB envelopes and entering the diffuse ISM. .  As mentioned earlier, this depletion factor applies to the PAHs ejected into the ISM from a post-AGB star.  Given the long lifetime of the ejected monomer PAHs into the ISM, $\sim$40 Myr, further depletion and cluster growth will occur through a variety of grain evolution, coagulation, and accretion processes \citep{jon19}.

\begin{center}
{\bf ORCHID iDs}
\end{center}
\noindent Lawrence S. Bernstein https:/orcid.org/0000-0001-6150-1579 \\
\noindent T. R. Geballe https://orcid.org/0000-0003-2824-3875


\end{document}